\pdfoutput=1

\documentclass[final,5p,11pt,authoryear,hidelinks,times]{elsarticle}

\usepackage[final]{microtype}

\usepackage{setspace}

\makeatletter
\let\@afterindenttrue\@afterindentfalse
\makeatother

\usepackage{titlesec}
\titleformat*{\section}{\Large\normalshape\bf}
\titleformat*{\subsection}{\large\normalshape\bf}
\titleformat*{\subsubsection}{\normalshape\bf}

\makeatletter
\renewcommand\paragraph{\@startsection{paragraph}{4}{\z@}%
  {-3.25ex\@plus -1ex \@minus -.2ex}%
  {1.5ex \@plus .2ex}%
  {\normalfont\normalsize\bfseries\itshape}}
\makeatother
\journal{NeuroImage}

\usepackage{hyperref}

\usepackage[font=normalsize,labelfont=bf]{caption}

\usepackage[utf8]{inputenc}

\usepackage{amsmath}
\usepackage{amsfonts}
\usepackage{bm}

\usepackage{tabularx}
\usepackage{booktabs}

\usepackage{xcolor}

\usepackage{graphicx}
\usepackage{subcaption}

\usepackage[status=final,index,multiuser,author=]{fixme}
\fxsetface{margin}{\linespread{1}\scriptsize}
\fxusetheme{color}

\usepackage[displaymath]{lineno}
\newcommand*\patchAmsMathEnvironmentForLineno[1]{%
  \expandafter\let\csname old#1\expandafter\endcsname\csname #1\endcsname
  \expandafter\let\csname oldend#1\expandafter\endcsname\csname end#1\endcsname
  \renewenvironment{#1}%
     {\linenomath\csname old#1\endcsname}%
     {\csname oldend#1\endcsname\endlinenomath}}%
\newcommand*\patchBothAmsMathEnvironmentsForLineno[1]{%
  \patchAmsMathEnvironmentForLineno{#1}%
  \patchAmsMathEnvironmentForLineno{#1*}}%
\AtBeginDocument{%
\patchBothAmsMathEnvironmentsForLineno{equation}%
\patchBothAmsMathEnvironmentsForLineno{align}%
\patchBothAmsMathEnvironmentsForLineno{flalign}%
\patchBothAmsMathEnvironmentsForLineno{alignat}%
\patchBothAmsMathEnvironmentsForLineno{gather}%
\patchBothAmsMathEnvironmentsForLineno{multline}%
}

\begin{document}

\begin{frontmatter}

  \title{
    Patient-specific solution of the electrocorticography\\
    forward problem in deforming brain}

\address[UWA]{
  Intelligent Systems for Medicine Laboratory,
  The University of Western Australia,\\
  35 Stirling Highway,
  Perth, WA, Australia}

\address[CRL]{
  Computational Radiology Laboratory,
  Boston Children's Hospital,
  Boston, MA, USA}

\address[Harvard]{
  Harvard Medical School,
  Boston, MA, USA}

\author[UWA]{B.F. Zwick\corref{mycorrespondingauthor}}
\ead{benjamin.zwick@uwa.edu.au}
\cortext[mycorrespondingauthor]{Corresponding author}

\author[UWA]{G.C. Bourantas}

\author[UWA]{S. Safdar}

\author[UWA]{G.R. Joldes}

\author[CRL,Harvard]{D.E. Hyde}

\author[CRL,Harvard]{S.K. Warfield}

\author[UWA]{A. Wittek}

\author[UWA,Harvard]{K. Miller}

\begin{abstract}

Invasive intracranial electroencephalography (iEEG) or electrocorticography (ECoG)
measures electric potential directly on the surface of the brain %
and can be used to inform treatment planning for epilepsy surgery.
Combined with numerical modeling
they can further improve accuracy of epilepsy surgery planning.
Accurate solution of the iEEG or ECoG forward problem,
which is a crucial prerequisite for solving the inverse problem
in epilepsy seizure onset zone localization,
requires accurate representation of the patient's brain geometry and tissue electrical conductivity
after implantation of electrodes.
However, implantation of subdural grid electrodes causes the brain to deform,
which invalidates preoperatively acquired image data.
Moreover, postoperative magnetic resonance imaging (MRI) is incompatible with implanted electrodes
and computed tomography (CT) has insufficient range of soft tissue contrast,
which precludes both MRI and CT from being used to obtain the deformed postoperative geometry.
In this paper, we present a biomechanics-based image warping procedure
using preoperative MRI for tissue classification
and postoperative CT for locating implanted electrodes
to perform non-rigid registration of the preoperative image data
to the postoperative configuration.
We solve the iEEG forward problem
on the predicted postoperative geometry
using the finite element method (FEM)
which accounts for patient-specific inhomogeneity and anisotropy of tissue conductivity.
Results for the simulation of a current source in the brain
show large differences in electric potential
predicted by the models based on the original images
and the deformed images
corresponding to the brain geometry
deformed by placement of invasive electrodes.
Computation of the lead field matrix
(useful for solution of the iEEG inverse problem)
also showed significant differences between the different models.
The results suggest that rapid and accurate solution of the forward problem
in a deformed brain for a given patient is achievable.

\end{abstract}

\begin{keyword}
epilepsy\sep
electroencephalography\sep
biomechanics\sep
diffusion tensor imaging\sep
meshless methods\sep
finite element method (FEM)
\end{keyword}

\end{frontmatter}

\section{Introduction}

Techniques to construct patient-specific models of brain bioelectric activity,
and to solve such models accurately and efficiently,
form a key enabling technology for neuroscience and neurology
\citep{
  baillet_etal_2001_electromagnetic,
  brette_destexhe_2012_handbook}.
Of particular interest is the application of such modeling and simulation techniques
to identification of epileptic seizure onset zones (SOZ)
which consider the brain at length scales
accessible by noninvasive neuroimaging
(such as CT, US and MRI),
and scalp electrodes, subdural electrode grids and strips, or depth electrodes
\citep{
  baillet_etal_2001_electromagnetic,
  brette_destexhe_2012_handbook}.
The spatial resolution of medical images such as MRI and CT,
and dimensions of scalp EEG and subdural ECoG electrodes---%
as well as the accuracy of surgery---%
are on the order of 1~mm,
but in current practice the SOZ is usually located with much lower accuracy
\citep{brodbeck_etal_2011_electroencephalographic,
  michel_brunet_2019_eeg,
  mouthaan_etal_2019_diagnostic}.

Epilepsy is a chronic brain disorder
that causes unpredictable and recurrent electrical activity in the brain (seizures).
Nowadays, epilepsy is treated with medication.
Although drugs do not cure epilepsy,
they can efficiently control seizures in up to 60\% of patients.
In the remaining 40\%, drug therapy fails to control seizures,
and surgical resection appears as a feasible alternative
\citep{engel_2003_greater,
  engel_2018_current,
  worldhealthorganization_2019_epilepsy}.
However, surgical treatment of epilepsy has been the most underutilized
of all proven effective therapeutic interventions in the field of medicine
\citep{engel_2003_greater,
  engel_2018_current,
  hader_etal_2013_complications,
  jette_wiebe_2015_health,
  jobst_cascino_2015_resective}.
This is because the success of surgical resection
depends on the accurate identification of the seizure onset zone (SOZ)
and it is often very difficult to characterize epileptic activity of the brain
and identify appropriate resection regions with sufficient accuracy to proceed with surgery.

\begin{figure*}
  \centering
  \includegraphics[width=\textwidth]{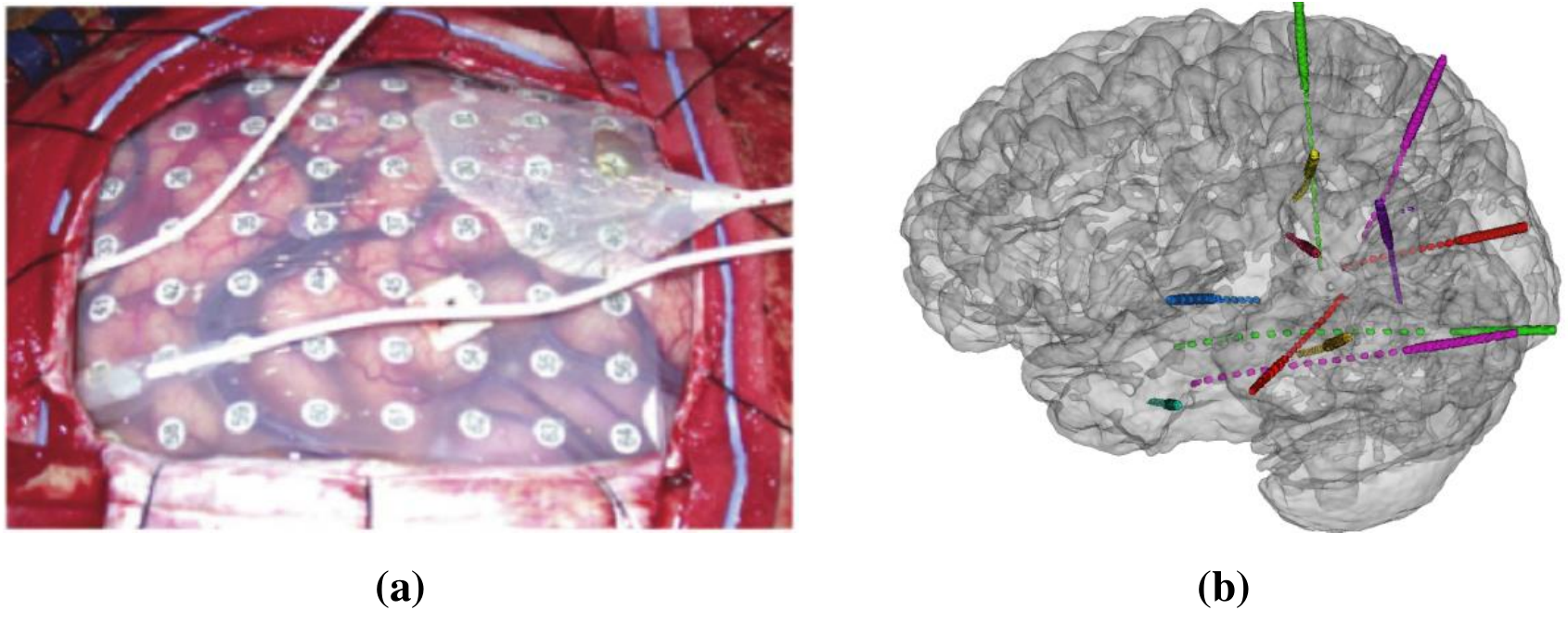}
  \caption{
    Electrodes used for recording electrical activity in the brain:
    (a) iEEG/ECoG mounted intracranial grid electrodes
    (modified image from \citet{husain_2015_practical}), and
    (b) SEEG electrodes implanted through holes in the skull
    (modified image from \citet{hyde_etal_2017_localization}).}
  \label{fig:brain-electrodes}
\end{figure*}

The SOZ is often identified with the help of intracranial measurement
of electric potential during a seizure.
An intracranial electroencephalogram (iEEG) or electrocorticography (ECoG)
\citep{jayakar_etal_2016_diagnostic}
records electrical activity from the cerebral cortex
using electrodes placed directly on the exposed surface of the brain
(Fig.~\ref{fig:brain-electrodes}a),
whereas stereoelectroencephalography (SEEG)
\citep{minotti_etal_2018_indications}
enables exploration of deeply located structures
using needle-like depth electrodes inserted inside the brain
(Fig.~\ref{fig:brain-electrodes}b).
Implantation of electrodes
is an invasive procedure and therefore applied rarely,
as the measurements by these electrodes
do not directly yield the location of the seizure onset zone
with accuracy comparable to imaging resolutions,
but rather form an input to further analysis.
Source modeling is one way to process that information
and combine it with other imaging information and prior assumptions.
To increase the utilization of epilepsy surgery
there is a pressing need for new and more effective methods
of SOZ localization that will enable clinicians to cure epilepsy
in a greater number of patients.
As it is estimated that 20 million people globally
have focal epilepsy that can be permanently cured by precisely targeted surgery
\citep{worldhealthorganization_2019_epilepsy},
methods for successful identification of the SOZ
would be of enormous health benefit.
One such promising technology is source localization from invasive measurements.
Using a volume conductor model of a patient's head
as the basis for an inverse problem,
source localization maps individual electrode measurements
to three dimensional maps of
distribution of electric potential within the brain.
These maps have the potential to significantly improve SOZ identification
by improved analysis of invasive EEG data.

Accurate and robust patient-specific models of brain bioelectric activity
could have applications well beyond guiding surgical treatment of epilepsy.
These include the analysis of extracranial measurement data,
such as standard electroencephalogram (EEG) and magnetoencephalogram (MEG)
often used in neurology, neuroscience and neuropsychiatry
\citep{brette_destexhe_2012_handbook},
as well as generation of high-quality synthetic training data
for artificial intelligence (AI) systems based on neural networks.
This last possible application is becoming particularly important,
as in recent years,
deep learning has started to gain popularity for bioelectric signal processing
\citep{
  cendes_mcdonald_2022_artificial,
  liu_etal_2021_probabilistic,
  sun_sclabassi_2000_forward,
  tobore_etal_2019_deep}.
With a large amount of data,
it outperforms traditional feature extraction in terms of classification accuracy
\citep{hu_etal_2019_mean}.
Deep learning algorithms, especially the convolutional neural network (CNN),
are also gradually being adopted in epilepsy management,
for example for seizure detection
\citep{gao_etal_2020_deep,thomas_etal_2018_eeg}.

Source localization based on ECoG or iEEG
can be split into two separate problems:
the forward problem that involves
calculating the electric potential within the brain volume
given a predefined source (or multiple sources); and
the inverse problem that involves
finding the source location (or locations)
given electric potential measurements at the sensor electrodes
\citep{baillet_etal_2001_electromagnetic,
  brette_destexhe_2012_handbook,
  hallez_etal_2007_review,
  grech_etal_2008_review}.
In addition to considerations specific to the inverse problem,
such as selection of the number of sources and
the choice of source localization algorithms
\citep{bradley_etal_2016_evaluation,grech_etal_2008_review,scherg_etal_2019_taking},
source localization requires
an accurate solution of the iEEG forward problem,
which depends on
an accurate representation of the geometry of the patient's brain
and electrical conductivity distribution
within the patients head
\citep{
  brette_destexhe_2012_handbook,
  hallez_etal_2007_review,
  vorwerk_etal_2014_guideline,
  wolters_etal_2006_influence}.

This study is based on the realization that a patient's brain
is significantly deformed by the implantation of the electrodes
(Fig.~\ref{fig:brain-electrodes}).
The shift in the brain surface caused by implantation of ECoG grid electrodes
is typically on the order of around 5~mm,
but often exceeds 10~mm
\citep{foldes_etal_2020_shift,hill_etal_2000_sources}.
Consequently, models built from preoperative images
that do not account for this brain shift
will be geometrically very inaccurate.
To remedy this situation,
we demonstrate how to construct a volume conductor model
corresponding to the true deformed brain geometry
and how to solve quickly the iEEG forward problem---%
a prerequisite for the iEEG inverse problem
that ultimately needs to be solved to localize sources.

In some previous studies,
to account for craniotomy-induced brain shift
during placement of grid electrodes,
the electrode positions identified from the postoperative CT were projected
onto the cortical surface extracted from the preoperative MRI
\citep{dykstra_etal_2012_individualized,
  yang_etal_2012_localization,
  tao_etal_2009_accuracy,
  laviolette_etal_2011_threedimensional,
  hermes_etal_2010_automated,
  taimouri_etal_2014_electrode}.
Although this approach solves the issue
of the electrodes being erroneously located within the brain tissue
instead of on the surface of the brain,
it does not accurately reflect the actual postoperative configuration of the brain.
In the current study,
instead of projecting the electrode positions onto the undeformed preoperative cortical surface,
we aim to predict the postoperative configuration of the brain
such that the cortical surface of the model used for solving the iEEG forward problem
corresponds to the actual position of the implanted surface electrode array.

In this paper,
we propose a novel modeling pipeline for iEEG source localization
that incorporates the tissue deformation
caused by the craniotomy and placement of the electrodes.
Using preoperative MR and postoperative CT images,
our approach employs biomechanics-based image warping
to predict the postoperative configuration of the brain
(i.e.\ the configuration deformed by the implantation of intracranial electrodes).
The postoperative CT is used to locate the electrode positions
and to inform the boundary conditions and loading of the biomechanical model.
The approach taken in this study
is to register the preoperative anatomical MRI and DTI onto the postoperative CT
(with electrodes implanted)
using a displacement field computed by a biomechanical model
\citep{miller_etal_2019_biomechanical_epilepsy,
  mostayed_etal_2013_biomechanical,
  safdar_etal_2021_automatic,
  wittek_etal_2007_patientspecific,
  wittek_miller_2020_computational,
  yu_etal_2022_automatic}
to obtain warped MRI and DTI
corresponding to the postoperative, deformed configuration of the patient's brain.
Based on these warped images we construct a geometrically correct computational domain
on which the partial differential equations of the iEEG forward problem are solved,
as well as inhomogeneous and anisotropic patient-specific distribution
of conductivity tensor used in the solution procedure.

Continuum models based on partial differential equations
are the dominant method for brain bioelectric activity modeling
at the spatial and temporal scales accessible via ECoG, iEEG, EEG or MEG.
Numerical solution of these equations
have attracted considerable attention in the literature.
Mainstream numerical methods, such as
finite element (FEM)
\citep{
  drechsler_etal_2009_full,
  marin_etal_1998_influence,
  medani_etal_2021_realistic,
  pursiainen_etal_2011_forward,
  schimpf_etal_2002_dipole,
  schrader_etal_2021_duneuro},
finite volume (FVM)
\citep{cook_koles_2006_highresolution},
finite difference (FDM)
\citep{bourantas_etal_2020_fluxconservative,
  hyde_etal_2012_anisotropic,
  saleheen_ng_1997_new,
  wendel_etal_2008_influence}
and boundary element methods (BEM)
\citep{acar_makeig_2010_neuroelectromagnetic,
  meijs_etal_1989_numerical,
  stenroos_sarvas_2012_bioelectromagnetic},
as well as newer, more sophisticated methods such as meshless methods
\citep{fietier_etal_2013_meshless}
that use a point cloud to represent the spatial domain,
have been used to numerically solve the governing equations
of the EEG forward problem.

The finite element method (FEM) achieves high numerical accuracy
\citep{
  drechsler_etal_2009_full,
  vorwerk_etal_2012_comparison,
  vorwerk_etal_2018_fieldtripsimbio}
and can be easily used to model complex geometries having anisotropic conductivities
\citep{
  gullmar_etal_2006_influence,
  gullmar_etal_2010_influence,
  haueisen_etal_2002_influence,
  rullmann_etal_2009_eeg,
  vorwerk_etal_2014_guideline,
  wolters_etal_2006_influence}.
The FEM uses a tessellation of the computational domain
in which the domain is partitioned into a set of elements of simple shape,
such as tetrahedrons or hexahedrons.
Tetrahedral meshes are generated by constrained Delaunay tetrahedralization
from segmented and reconstructed tissue surface representations
\citep{vorwerk_etal_2014_guideline}.
On the other hand, regular hexahedral meshes can be easily generated
from voxel-based raw image data
\citep{schimpf_etal_2002_dipole,rullmann_etal_2009_eeg,vorwerk_etal_2017_mixed}.
This greatly simpliﬁes mesh generation
which is an important consideration
for ensuring compatibility with clinical workflows.
Moreover, regular hexahedra have favourable numerical properties
over skewed hexahedra or tetrahedra
\citep{hughes_2000_finite},
Geometry-adapted hexahedra
\citep{wolters_etal_2007_geometryadapted}
or tetrahedra
\citep{vorwerk_etal_2017_mixed}
may give similar results using less elements
but this comes at the significant cost of
additional complications of mesh generation.

Our methodology is an embodiment of the ``image-as-a-model'' concept
\citep{zhang_etal_2013_patientspecific}
as it is entirely voxel-based and the finite element method
with a mesh coinciding with the voxelized structure of the image
is used to obtain the solution.
The advantage of this voxel-based analysis includes
straight-forward generation of realistic patient-specific models
from three-dimensional image data
and a simple data structure
that is suitable for storage, handling, numerical solution and visualization
\citep{rullmann_etal_2009_eeg,
  schimpf_etal_1998_realistic,
  schimpf_etal_2002_dipole}.

To demonstrate the application of the proposed methodology
and to quantify the effect of geometric accuracy
on the model solutions,
we solve two example problems
using the original preoperative geometry
(with both the actual and projected electrode positions)
and the geometry after model based image deformation.
In the first example,
we apply a current dipole inside the brain
and compare the potential computed at the electrodes.
In the second example,
we compute the lead field matrices
(gain matrices relating the source space with the measurement space %
which are often used in the solution of the EEG inverse problem)
using the models constructed from
the original and deformed image data,
and compare the difference between the two sets of results.
In both examples,
the large difference in the results
obtained using the original and deformed image data
suggest that the forward problem solution accuracy
may be improved by accounting for electrode-induced brain shift
in the patient-specific volume conductor model
as described herein,
which,
when combined with suitable inverse solution methods,
may lead to improved accuracy in source localization.

The paper is organized as follows.
In section 2,
we describe the mathematical model,
patient-specific model construction
and the components of the numerical methodology.
In section 3,
we provide patient-specific numerical results
confirming robustness and efficiency of our modeling and simulation pipeline,
as well as demonstrating inaccuracies of using models
built directly from preoperative, undeformed images.
Section 4 contains discussion and conclusions.

\section{Methods}

\subsection{Governing equations of the brain bioelectric activity at scales as measured by EEG and MRI}

The electroencephalography (EEG) forward problem
involves predicting the electric potential within the brain
and at the implanted electrodes
given a predefined source.
The relevant frequency spectrum in EEG is typically below 1 kHz,
with most studies dealing with frequencies between 0.1 and 100 Hz
\citep{baillet_etal_2001_electromagnetic,brette_destexhe_2012_handbook}.
Therefore, the physics of EEG can be approximated by Poisson's equation,
which is the quasi-static approximation of Maxwell's equations.
For a spatial domain \(\Omega \in \mathbb{R}^3\)
with boundary \(\partial \Omega = \overline{\Gamma_D \cup \Gamma_N}\)
and outward unit normal \(\bm{n}\),
Poisson's equation for the EEG forward problem can be written as follows:
\begin{align}
  \label{eq:poisson-eq}
  -\nabla \cdot (C (\nabla u)) &= f \text{ in } \Omega, \\
  \label{eq:poisson-nbc}
  \bm{n} \cdot (C (\nabla u))  &= g \text{ on } \Gamma_N, \\
  \label{eq:poisson-dbc}
  u &= h \text{ on } \Gamma_D,
\end{align}
where
$u$ is the unknown scalar potential and
$C$ is the (symmetric positive semi-definite) conductivity tensor.
The low conductivity of air outside the scalp
(\(C = 0\) for all \(x \notin \bar\Omega\))
implies that a zero-flux Neumann boundary condition $g=0$
can be applied on the surface $\Gamma_N$.
Dirichlet boundary conditions \(h\) on the surface $\Gamma_D$ are typically applied
by setting the potential to zero at a node corresponding to a reference electrode.
Loading can be applied to the model either by prescribing the potential \(u\)
at certain nodes or by applying a current source
\(f = \nabla \cdot \bm{j}\),
with \(\bm{j}\) being a dipole source vector.
The epileptic seizure onset source is typically modeled as a current dipole
\citep{hallez_etal_2007_review}.

Electromagnetic source localization employs a linear model,
known as the lead field matrix,
which relates the measured electrode voltages
to the cerebral current sources that generated them
\citep{rush_driscoll_1969_eeg,weinstein_etal_2000_leadfield}.
The lead field matrix can be computed using
Helmholtz's principle of reciprocity,
which relates the potential difference between two points \(A\) and \(B\)
caused by a given dipole \(\bm{p}\),
to the electric field \(\bm{E} = -\nabla u\)
at the dipole location resulting from a current \(I\)
placed between \(A\) and \(B\)
\citep{plonsey_1963_reciprocity,weinstein_etal_2000_leadfield}:
\begin{equation}
  \label{eq:reciprocity}
  \frac{\bm{E}\cdot\bm{p}}{-I} = u_A - u_B.
\end{equation}
The lead field matrix can be constructed
by placing a source and sink at pairs of electrodes,
and computing for each of these pairs
the resulting electric field in all of the elements.
The reciprocity principle can then be applied
to reconstruct the potential differences at the electrodes
for a source placed in any element
\citep{weinstein_etal_2000_leadfield}.

Computation of the lead field matrix requires the solution
of the forward problem for each sensor and ground electrode pair.
To compute the lead field matrix,
one electrode is selected arbitrarily as the ground (or reference) electrode.
For each lead field (row in the lead field matrix),
a unit current is applied to one of the sensor (non-ground) electrodes.
The electric field in each element
is computed as the gradient of the electric potential
and represents a row in the matrix.
The potential at the \((M-1)\) non-ground electrodes
for a source located in any of the \(N\) elements
is given by
\begin{equation}
  \bm{u} = \bm{L} \bm{j},
\end{equation}
where
\(\bm{L}\), of dimension \( (M-1) \times 3N \),
is the lead field matrix, and
\(\bm{j}\), of dimension \(3N \times 1\),
is a vector of dipole amplitudes in all points of the source space.

Although the iEEG forward problem is linear,
it is difficult to solve due to
significant difficulties in creating a patient-specific domain \(\Omega\)
  on which the equations \eqref{eq:poisson-eq}--\eqref{eq:poisson-dbc} are to be solved
  and patient-specific spatial distribution of anisotropic conductivity tensor \(C\),
  addressed below.

\subsection{Patient-specific model generation}

\begin{figure*}
  \centering
  \includegraphics[width=0.8\linewidth]{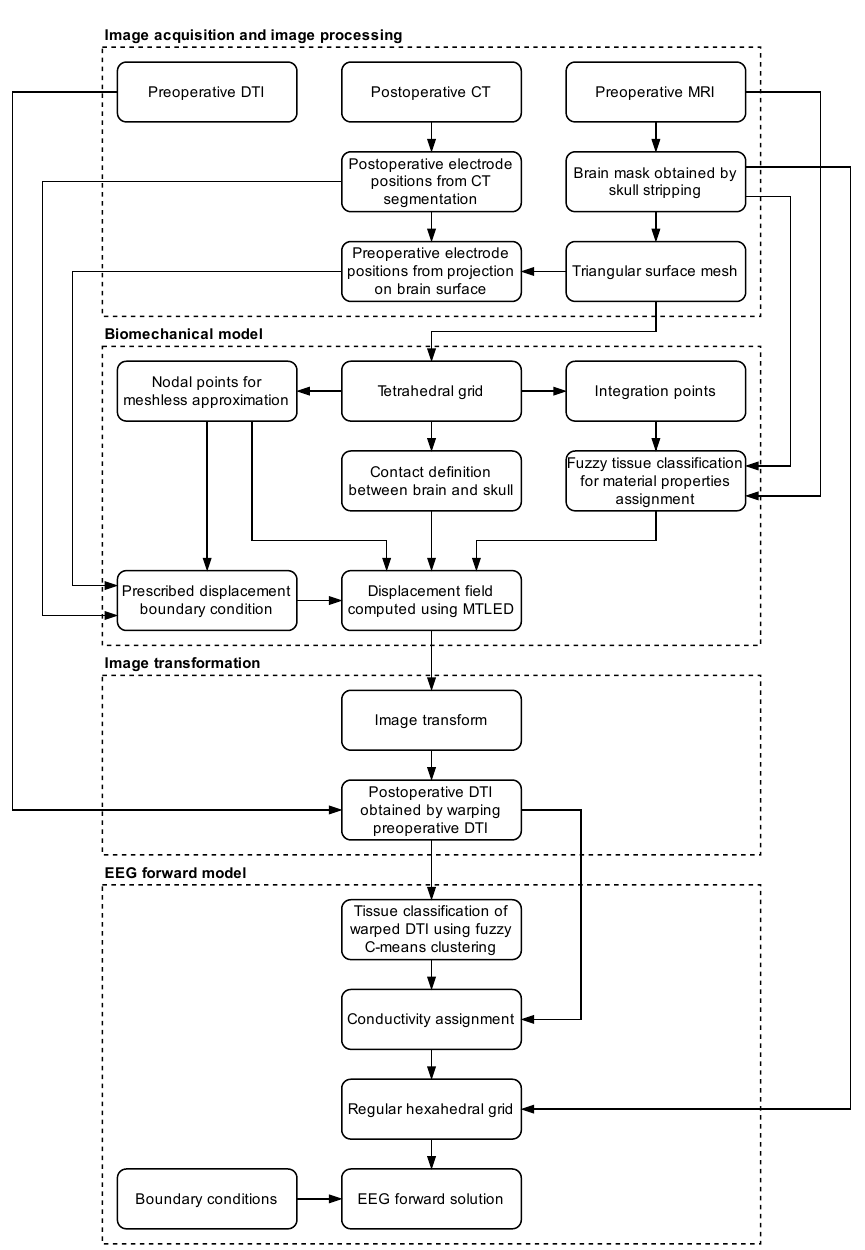}
  \caption{Flowchart of the patient-specific solution of the EEG forward problem in deforming brain.
    Brain-shift caused by implantation of electrodes is computed using the biomechanical model.
    The computed displacement field is used to transform the DTI to the postoperative configuration.
    This warped DTI is then used as the basis for creating the EEG forward model.}
  \label{fig:flowchart}
\end{figure*}

In this section,
we describe the methods used to create a detailed patient-specific model
for the EEG forward problem.
Accurate solution of the EEG forward problem
requires a correspondingly accurate representation of both
the computational domain
(\(\Omega\) in Eq.~\eqref{eq:poisson-eq})
and the tissue conductivity
(\(C\) in Eq.~\eqref{eq:poisson-eq})
throughout the conducting head volume.
The head volume is bounded by the rigid skull and scalp
which can be easily extracted from the preoperative images.
The brain, however,
is a soft tissue
that deforms due to the insertion of the electrode grid array.
To take into account the brain shift following electrode insertion
we apply our biomechanics-based image warping techniques
to predict the postoperative brain geometry.
Fig.~\ref{fig:flowchart}
provides a flowchart of the proposed methodology.
A detailed description of each step
is provided in the following subsections.

We created the following three models for comparison
of the different modeling approaches:
\begin{enumerate}
\item \emph{unwarped}
  model based on preoperative (unwarped) image data
  with actual electrode positions identified from postoperative CT;
\item \emph{projected}
  model based on preoperative (unwarped) image data
  with electrode positions projected
  from their actual locations identified from postoperative CT
  onto the cortical surface; and
\item \emph{warped}
  model based on postoperative (warped) image data
  predicted by biomechanics-based image registration
  with actual electrode positions identified from postoperative CT.
\end{enumerate}

\subsubsection{Image acquisition and image processing}
\label{sec:image-processing}

To create the patient-specific EEG forward model we require
preoperative anatomical magnetic resonance image (MRI)
to obtain the undeformed brain geometry;
preoperative diffusion tensor image (DTI)
to classify tissue types at each voxel and
to estimate patient-specific tissue conductivity;
and
postoperative computed tomography (CT)
to locate the implanted electrodes.

\begin{figure*}
    \centering
    \includegraphics[width=\textwidth]{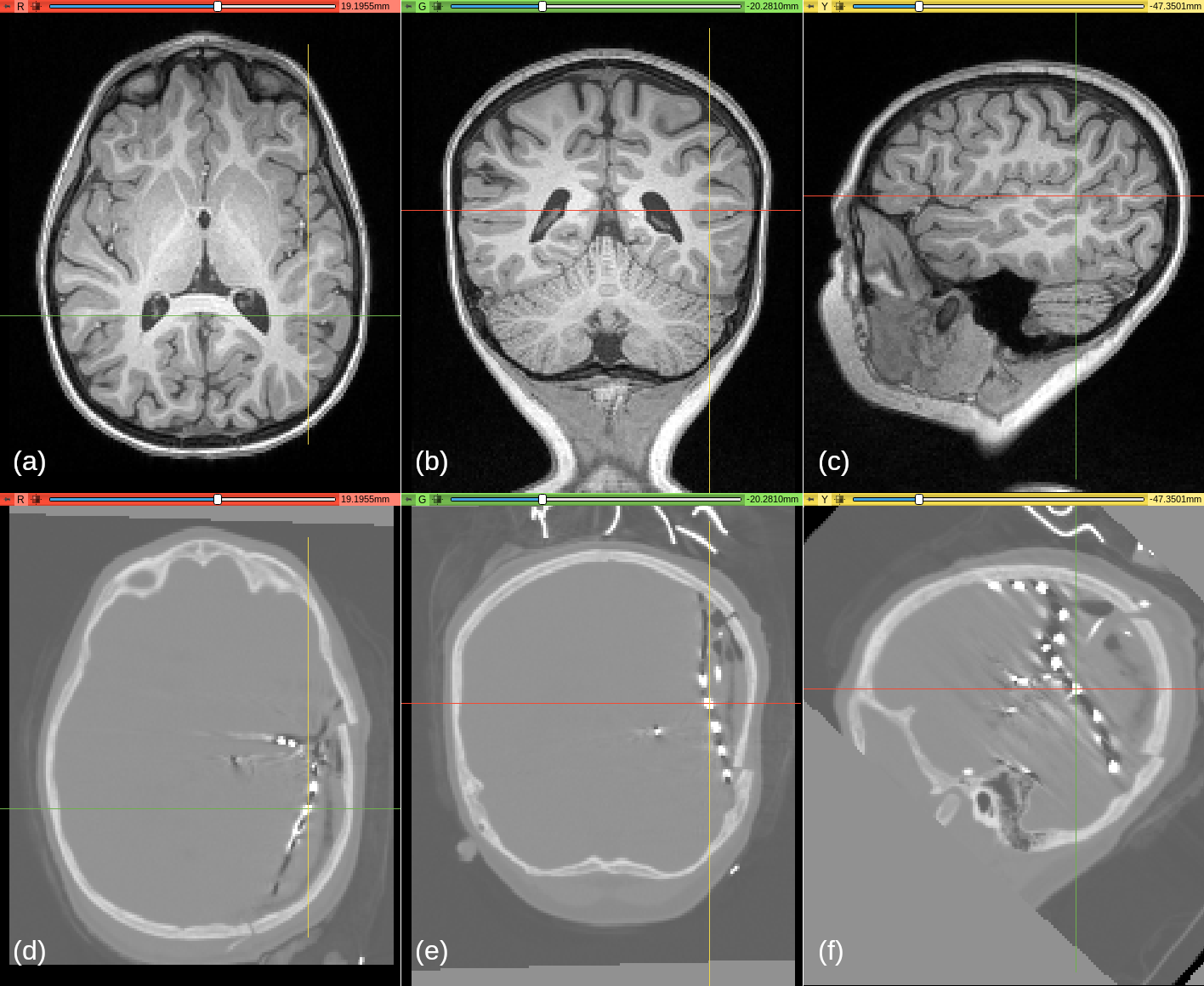}
    \caption{
    Axial, coronal and sagittal slices of the
    (a,b,c) preoperative MRI and
    (d,e,f) postoperative CT
    rigidly aligned with the preoperative MRI.
    }
    \label{fig:image-data}
\end{figure*}

T1 and T2 weighted structural MRI scans %
and diffusion MRI %
were collected from
a 12 year old female epilepsy patient
under evaluation for surgical intervention
at Boston Children's Hospital
(BCH ethics approval no.\ IRB-P00025254,
UWA ethics approval no.\ RA/4/1/9336).
The T1w MPRAGE image was acquired in the saggital plane
at a nominal resolution of \(1 \times 1 \times 1\)~mm
(21~cm FOV, inplane matrix size of \(192 \times 192\) and 160 slices).
The T2w TSE MRI was acquired at nominal resolution of \(0.4 \times 0.4 \times 2.4\)~mm.
The DWI acquisition was at a nominal resolution of \(1.7 \times 1.7 \times 2.0\)~mm.
All images were coregistered to the T1 scan using a rigid transformation,
estimated by optimizing the mutual information between the two images
\citep{grau_etal_2004_improved,weisenfeld_warﬁeld_2009_automatic}
as implemented in CRKIT
(\url{http://crl.med.harvard.edu/software}),
and resampled to have matching 1~mm isotropic resolution.
Residual distortion and patient motion in the DWI
was compensated for by alignment to the T1w MPRAGE
and appropriate reorientation of gradient directions
\citep{peters_etal_2012_loss,ruiz-alzola_etal_2002_nonrigid}.
Diffusion tensors were estimated using robust least squares.
The dimensions of the images are $160 \times 192 \times 192$ voxels
with spacing \(1 \times 1.09375~\mathrm{mm} \times 1.09375~\mathrm{mm}\)
(Fig.~\ref{fig:image-data}).

An \(8\times8\) grid of platinum-iridium electrode disks
(4~mm diameter, 10~mm spacing)
embedded in a non-conductive silastic substrate
(2~mm thickness)
was placed on the cortical surface through a craniotomy.
Electrode placements were identiﬁed from postoperative CT,
coregistered and resampled to the resolution of the preoperative MRI.
Intensity thresholding was used to identify individual electrodes,
with manual validation of electrode separation.

The postoperative electrode positions
were projected onto the preoperative brain surface
to obtain the displacement of the brain surface
in the vicinity of the electrodes
that is used to define loading in the biomechanical model
(see section~\ref{sec:patient-specific-geometry}),
and to correct for the intraoperative brain shift
in the model that was based on undeformed preoperative image data
with projected electrode positions
\citep{taimouri_etal_2014_electrode}.
We used the ImplicitPolyDataDistance method
of the Visualization Toolkit
\citep{schroeder_etal_2006_visualization}
to compute the distance from each electrode as seen on CT
to the corresponding nearest point on the undeformed brain surface mesh.

\subsubsection{Patient-specific geometry}
\label{sec:patient-specific-geometry}

\paragraph{Brain deformation due to intracranial electrode array insertion}

Insertion of intracranial electrodes
results in significant brain deformations
that need to be accounted for when defining the domain
on which the EEG forward problem needs to be solved.
The deformed (with respect to the original preoperative MRI and DTI)
postoperative configuration of the brain was obtained
using biomechanics-based image warping
\citep{miller_etal_2019_biomechanical_epilepsy}.
In this approach,
the brain tissue is modeled as a deformable solid
and the displacement field is computed using
a meshless total Lagrangian explicit dynamics algorithm
\citep{joldes_etal_2019_suite}.
The methodology used to compute the brain shift,
caused in this study by the implantation of electrodes,
has been extensively validated %
in our previous studies
\citep{
  garlapati_etal_2014_more,
  ma_etal_2011_effects,
  miller_etal_2011_computational,
  miller_etal_2019_biomechanical,
  mostayed_etal_2013_biomechanical,
  wittek_etal_2007_patientspecific,
  wittek_etal_2010_patientspecific,
  wittek_miller_2020_computational}.

The computational grid for the biomechanical model was created as follows.
We applied the skull stripping procedure available in
FreeSurfer (\url{http://surfer.nmr.mgh.harvard.edu})
\citep{dale_etal_1999_cortical},
an open-source software suite for processing and analyzing
human brain MRIs,
to the preoperative T1-weighted anatomical MRI to create a brain mask
(i.e., a binary label map that is nonzero in the region of the brain only).
The triangulated surface of the brain mask segment
was extracted and remeshed using PyACVD
(\url{https://github.com/pyvista/pyacvd})
to obtain a high quality uniformly refined surface mesh.
A tetrahedral grid was generated from the surface mesh of the brain using Gmsh
(\url{https://gmsh.info})
\citep{geuzaine_remacle_2009_gmsh}.
The vertices of this grid
were used as nodes for the meshless approximation
of the variable of interest (displacement),
while the tetrahedral cells
were used for numerical integration of the weak form.
The meshless biomechanical model contained
18,434 nodes (includes rigid skull nodes),
44,821 tetrahedral integration cells, and
179,284 integration points.
Note that the computational methods and grid
used to solve the biomechanical model
as described in this section
are different to those used for the EEG forward problem,
which is solved using the finite element method with a regular hexahedral grid
as described in section~\ref{sec:solution-eeg}.

The mechanical response of brain tissue
was modeled using a modified neo-Hookean constitutive equation
with strain energy density
\citep{zienkiewicz_etal_2013_finite_solid}
\begin{equation}
  \label{eq:consitutive-neohookean-zienkiewicz}
  W(I_1, J) =
  \frac{\mu}{2} \left( (J)^{-2/3} I_1 - 3 \right) +
  \frac{\kappa}{2} (J - 1)^2,
\end{equation}
where
\(J\) is the determinant of the deformation gradient and
\(I_1\) is the first invariant of the right Cauchy-Green deformation tensor.
The shear modulus, \(\mu\), and Lam\'e's first parameter, \(\lambda\),
are related to the Young's modulus, \(E\), and Poisson's ratio, \(\nu\),
by the following relationships:
\begin{equation}
  \mu = \frac{E}{2(1 + \nu)}, \qquad
  \lambda = \frac{E\nu}{(1+\nu)(1-2\nu)}.
\end{equation}
The material parameters $E$ and $\nu$
were assigned to each voxel using fuzzy tissue classification
\citep{li_etal_2016_fuzzy,zhang_etal_2013_patientspecific}
conducted on preoperative MRI.
As in our previous work
\citep{wittek_etal_2010_patientspecific,
  miller_etal_2019_biomechanical_epilepsy},
we used
$E=3000$~Pa and $\nu=0.49$ for brain tissue, and
$E=100$~Pa and $\nu=0.1$ for cerebrospinal fluid.

As the brain deformations are caused by electrode array insertion,
the loading is defined based on the displacement
between the brain surface of the preoperative MRI
and the location of electrodes in the postoperative CT,
rigidly registered to preoperative MRI.
The displacement magnitude at each node was obtained by
first projecting the electrode centroids from the postoperative CT
onto the preoperative brain surface.
The displacements of the grid nodes were obtained
by interpolating the displacements from the electrode centroids
using moving least squares (MLS) approximation
\citep{lancaster_salkauskas_1981_surfaces}.
The maximum displacement applied was 21.9~mm.
The brain deformation problem with displacement loading
is a Dirichlet-type problem
and therefore the computed displacements are only weakly sensitive
to the assumed material model
\citep{wittek_etal_2009_unimportance,
  ma_etal_2011_effects,
  miller_lu_2013_prospect}.

The skull was modeled as a rigid surface %
approximated
by the outer surface of the brain mask.
Contact between the brain and the skull was modeled using
a frictionless finite sliding contact algorithm
\citep{joldes_etal_2008_realistic}.

The biomechanical model was solved using the
meshless total Lagrangian explicit dynamics (MTLED) algorithm
\citep{horton_etal_2010_meshless,
  miller_etal_2012_finite,
  joldes_etal_2019_suite}.
The MTLED method uses modified moving least squares (MMLS) shape functions
\citep{joldes_etal_2015_modified}
and an explicit central difference method
with adaptive dynamic relaxation to obtain the static solution
\citep{joldes_etal_2011_adaptive}.

\paragraph{Image transformations using the computed displacement field}

We projected the displacement vector field onto the image grid,
using the same MMLS shape functions
as those used to obtain the solution of the biomechanical model,
to obtain the forward displacement field transform.
The forward displacement field transform maps each point to its transformed position.
However, for image warping the inverse transform
that maps each point to its original location is required.
The forward displacement field transform was inverted using 3D~Slicer
(\url{https://www.slicer.org})
\citep{fedorov_etal_2012_3d},
allowing it to be applied directly to scalar, vector or tensor images.

\begin{figure*}
    \centering
    \includegraphics[width=\textwidth]{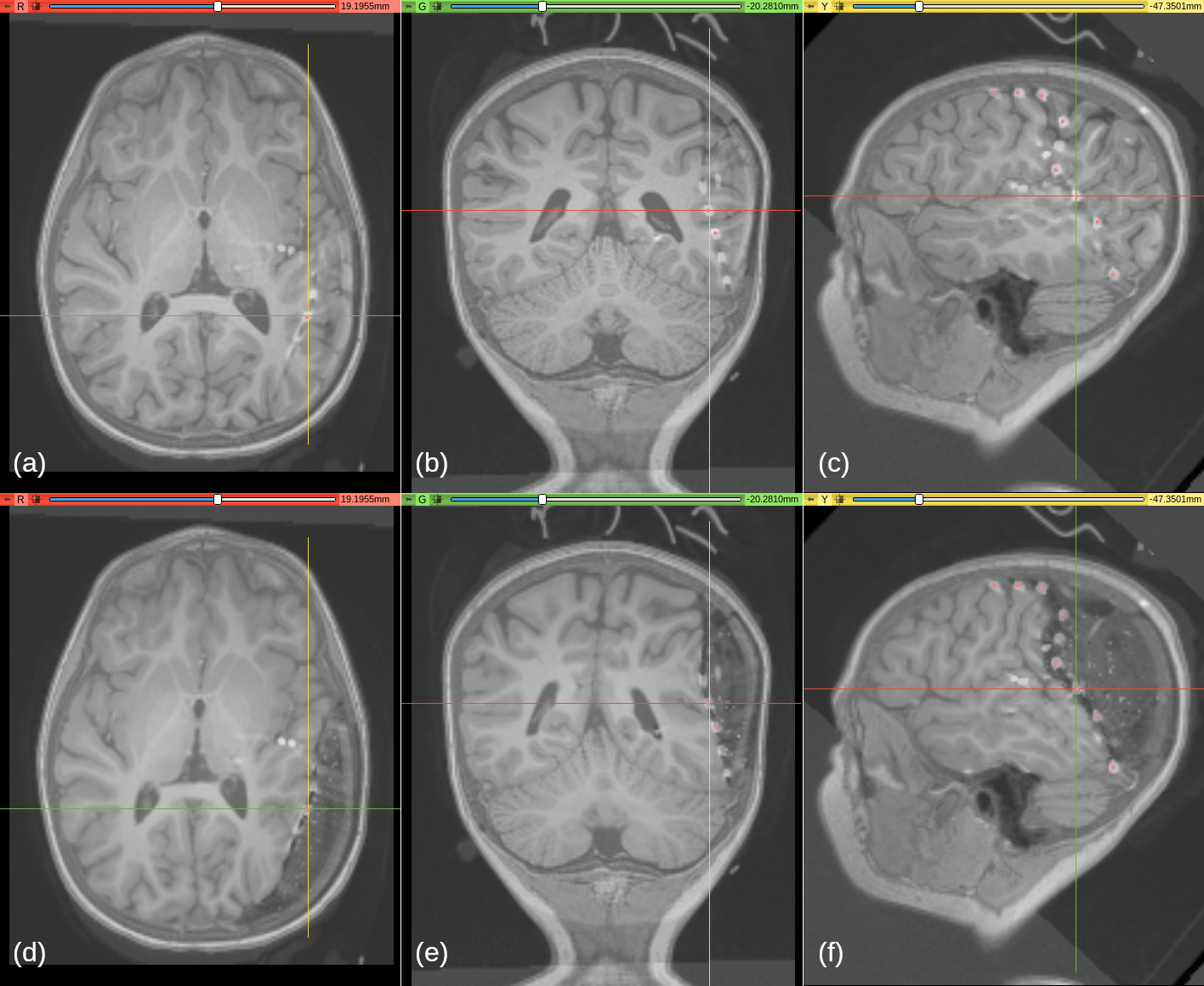}
    \caption{
    Original (actual preoperative) and deformed (predicted postoperative) MR images compared
    with original CT image and electrode positions.
    Postoperative CT image and electrode positions
    (white spheres in CT and red points in the slice planes)
    are overlayed on the
    (a,b,c) MRI acquired preoperatively and
    (d,e,f) MRI registered to postoperative configuration of the brain
    obtained using biomechanics-based image warping.
    }
    \label{fig:mri-ct-elec}
\end{figure*}

The scalar images (MRIs)
were transformed using the ResampleScalarVectorDWIVolume module
in 3D~Slicer.
Fig.~\ref{fig:mri-ct-elec}
shows the electrode positions
with respect to the original and deformed image data.

\begin{figure*}
    \centering
    \includegraphics[width=\textwidth]{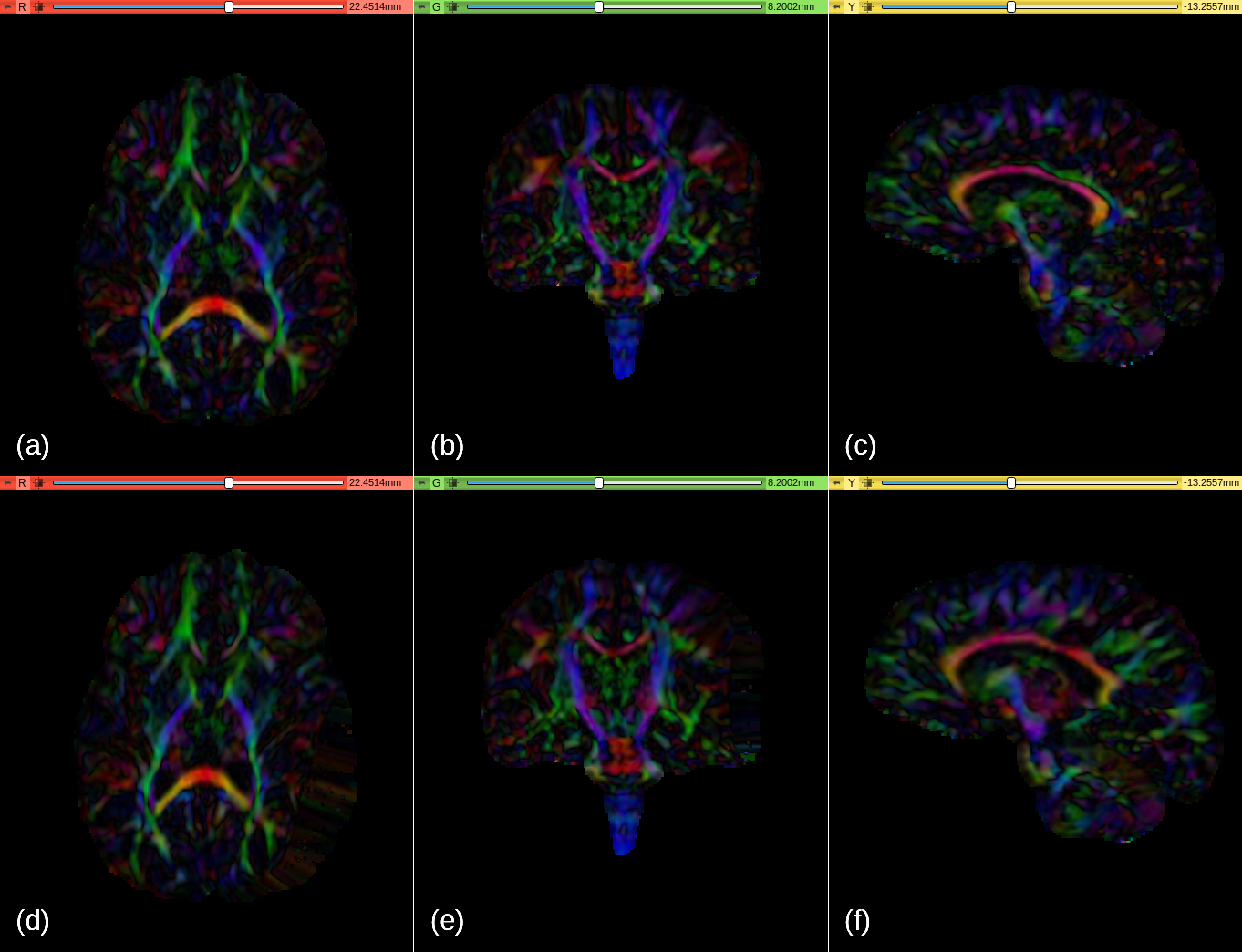}
    \caption{
    Diffusion tensor images (DTIs) of the brain
    (a,b,c) acquired preoperatively, and
    (d,e,f) registered to postoperative configuration of the brain
    using biomechanics-based image warping.
    Fiber orientation is denoted by
    red (left--right orientation),
    green (anterior--posterior orientation) and
    blue (superior--inferior orientation)
    colors.
    }
    \label{fig:dti-orientation}
\end{figure*}

The diffusion tensor image (DTI) was transformed using the
ResampleDTIVolume module in 3D~Slicer
with linear interpolation
and the preservation of the principal direction (PPD)
tensor transformation method
\citep{alexander_etal_2001_spatial}.
Fig.~\ref{fig:dti-orientation}
shows the orientation of the diffusion tensors
with respect to the original and deformed image data.

\subsubsection{Patient-specific conductivity tensor distribution}
\label{sec:dti-segmentation}

\paragraph{Voxel labeling}

In order to assign conductivity tensors to the image voxels
we consider five tissue types in the image:
skull, scalp, white matter (WM), gray matter (GM)
and cerebrospinal fluid (CSF).
Scalar quantities derived from the diffusion tensor,
such as mean diffusivity and fractional anisotropy,
can be used to classify brain tissue
into CSF, GM and WM %
based on their diffusion properties
\citep{pierpaoli_etal_1996_diffusion}.
For the purpose of this study,
we performed the tissue classification of the original and deformed DTIs using fuzzy C-means clustering
in two steps.
In the first step,
we used the mean diffusivity
to separate the CSF from the brain tissue.
In the second step,
we used the fractional anisotropy
to separate the WM from the GM.
The fuzziness parameter was set to \(m=2\).
We do not generate any segmentation surfaces %
and we do not require that
parts %
of the same tissue class are connected.
This method of tissue classification does not require any user interaction
which greatly simplifies the patient-specific model generation.

\begin{figure*}
  \centering
  \includegraphics[width=\textwidth]{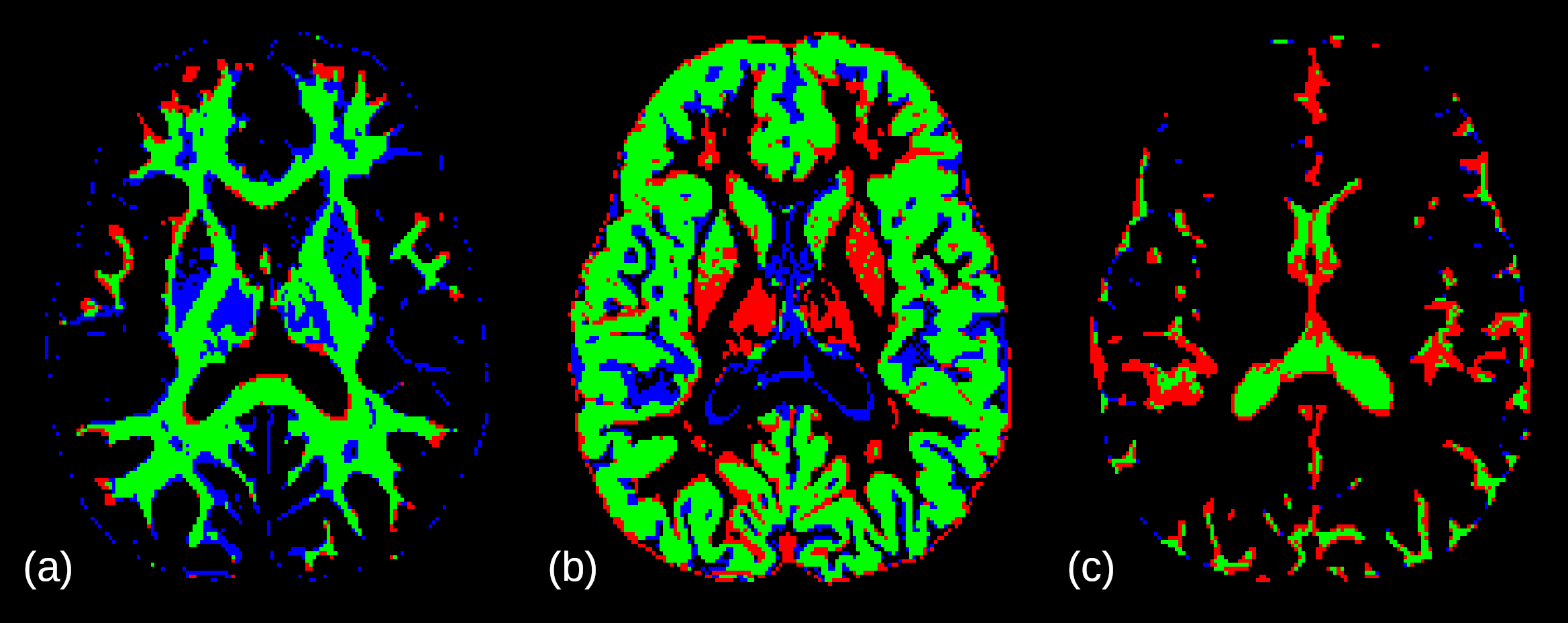}
  \caption{
    Axial slice of the brain showing
    voxels classified differently as
    (a) white matter,
    (b) gray matter or
    (c) cerebrospinal fluid (CSF)
    using the STAPLE method (blue)
    and our approach (red).
    Voxels that were classified as the same tissue class using both methods
    are shown in green.
  }
  \label{fig:staple-figure}
\end{figure*}

To evaluate the accuracy of our automated DTI-based tissue classification procedure
we compared our segmentation results with those obtained using the STAPLE method
\citep{warfield_etal_2004_simultaneous}.
Table~\ref{tab:staple-table}
lists the number of voxels that belong to white matter, gray matter and CSF
using the different segmentation methods.
The results are very close as shown in Fig.~\ref{fig:staple-figure}.

\begin{table*}[t]
  \caption{Number of white matter (WM), gray matter (GM) and cerebrospinal fluid (CSF)
    voxels using the current voxel classification method and STAPLE.
  }
  \label{tab:staple-table}
  \centering
\begin{tabularx}{\textwidth}{@{}Xcccccc@{}}
\toprule
& \multicolumn{3}{c}{STAPLE} & \multicolumn{3}{c}{Our method}\\
\cmidrule{2-4} \cmidrule{5-7}
 & Voxels & Volume (cm$^3$) & Percentage & Voxels & Volume (cm$^3$) & Percentage\\
\midrule
WM    &   291,185 &     348.3 &      28.4 &   214,381 &     256.5 &      20.3 \\
GM    &   660,152 &     789.7 &      64.5 &   673,756 &     806.0 &      63.7 \\
CSF   &    72,434 &      86.7 &       7.1 &   169,119 &     202.3 &      16.0 \\
Total & 1,023,771 &    1224.7 &     100.0 & 1,057,256 &    1264.8 &     100.0 \\
\bottomrule
\end{tabularx}
\end{table*}

More sophisticated methods,
some also based on fuzzy C-means clustering
\citep{wen_etal_2013_brain},
may be used to improve the labeling accuracy.
Segmentation using automated methods that use atlas-based approaches, however,
may be difficult to apply because the postoperative configuration of the brain
is often very different to a typical healthy brain. %
The relatively simple automated approach used here
demonstrates that the method may be potentially applied in the clinic
with minimal manual intervention.

\paragraph{Fusion of preoperative MRI and DTI, postoperative CT, and deformed MRI and DTI data}

\begin{figure*}
    \centering
    \includegraphics[width=\textwidth]{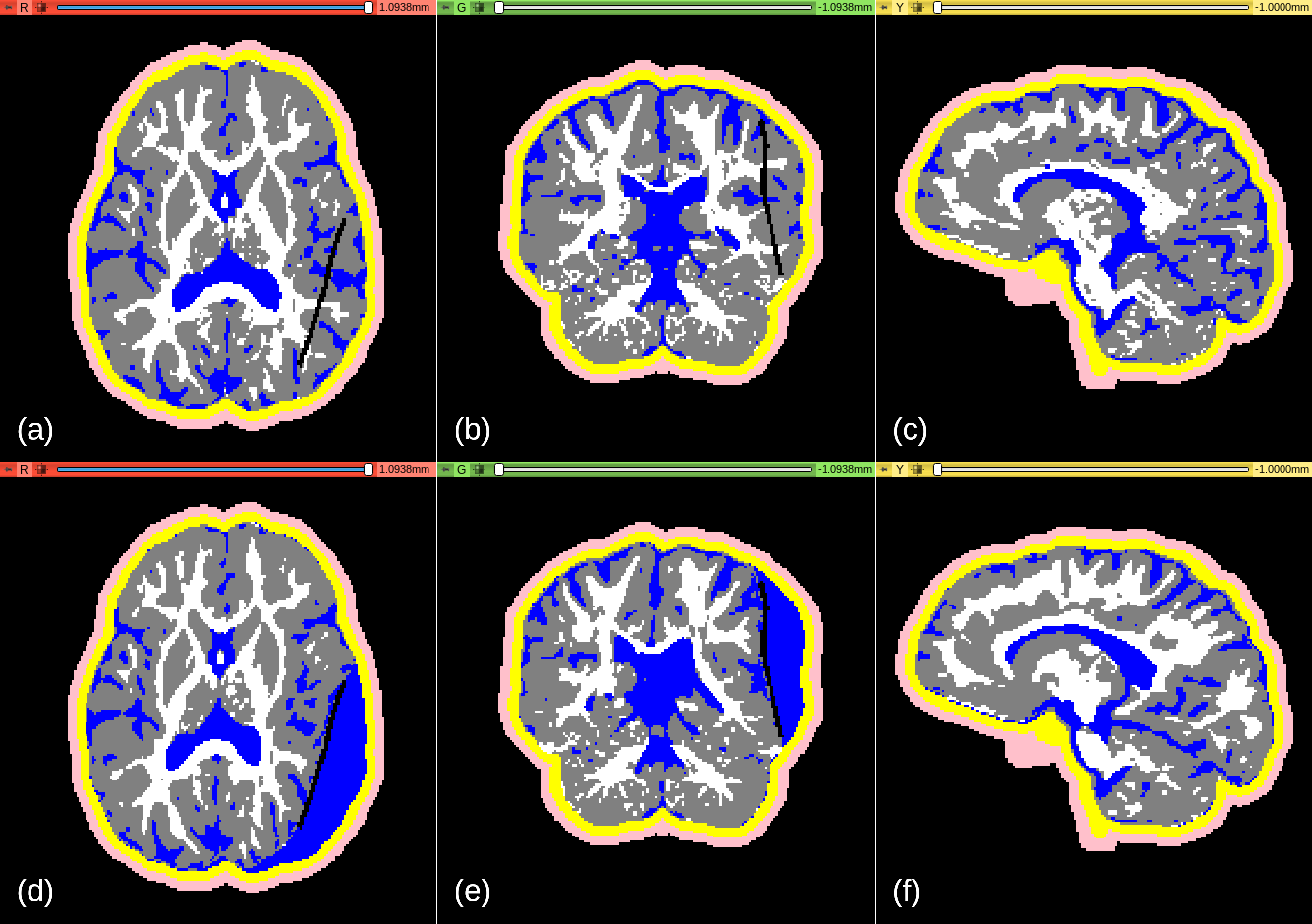}
    \caption{
    Tissue label maps based on
    (a,b,c) original preoperative and
    (d,e,f) deformed by insertion of electrodes postoperative
    image data.
    Tissue classes are colored as follows:
    scalp (pink); skull (yellow); GM (gray); WM (white); and CSF (blue).
    The location of the electrode grid array can be identified by the line of black voxels
    in the vicinity of the right temporal and parietal lobes.
    }
    \label{fig:brain-labelmaps}
\end{figure*}

Two label maps,
one based on the original preoperative MRI
and the other based on the warped MRI
corresponding to the postoperative configuration of the brain were created
by combining the original preoperative data
(skull and scalp)
with the original preoperative or predicted postoperative data
(original or deformed brain classified as WM, GM or CSF)
and actual postoperative CT (electrode grid array).
The cavity between the brain and the skull
was filled with CSF.
Fig~\ref{fig:brain-labelmaps} shows the label maps for the EEG forward models
constructed using the original preoperative and
deformed by electrode insertion postoperative MR images.
In the preoperative image,
the region between the skull
and the ECoG electrodes that were placed on the brain surface
is labelled as brain tissue
but this clearly does not match the post-implantation situation.
This indicates that using preoperative, undeformed images
for patient-specific geometry generation
may yield very inaccurate models.

The skull is difficult to segment from the CT image
because electrodes and beam hardening artifacts pollute the image
with regions of similar intensity as the skull.
Many of these regions overlap with bone tissue of the skull
which makes it impossible to apply simple thresholding methods
to extract the bone tissue.
Furthermore, fusion of the deformable soft tissue segments
and the rigid but resected and non-conforming skull segment
would add significant complications to the imaging pipeline.
Therefore, for simplicity,
we included the skull and scalp regions in the model
by offsetting the brain surface
by 4.4~mm (4 voxels).
The effect of this simplified skull and scalp geometry
on the model results should be negligible.
For invasive iEEG or ECoG, as compared to non-invasive scalp EEG,
the measurements are taken directly from the brain surface
and attenuation by the skull is minimal
which suggests that the electrical properties of the skull
have little effect on the electric field within the brain
\citep{hallez_etal_2007_review}.
Alternatively, the skull may be segmented from the CT image
to provide a more realistic representation of the geometry,
but this requires time consuming manual segmentation
which is incompatible with a clinical workflow.

The voxels corresponding to the \emph{actual} and \emph{projected} locations
of the electrode grid array substrate
were identified by fitting a surface
through the \emph{actual} electrode centroids visible in the CT image
and the electrode centroids \emph{projected} onto the cortical surface
as described in section~\ref{sec:image-processing}.
The electrode centroids were triangulated to generate a surface
representing the electrode grid array substrate.
This surface was refined and extruded by the equivalent distance of 2 voxels
in the direction away from the brain to create a closed volume.
The volume was used to create a segment representing the substrate
that is at least one layer thick throughout
to ensure that there was no current leakage.

\subsubsection{Conductivity tensor assignment}

\begin{figure*}
    \centering
    \includegraphics[width=\textwidth]{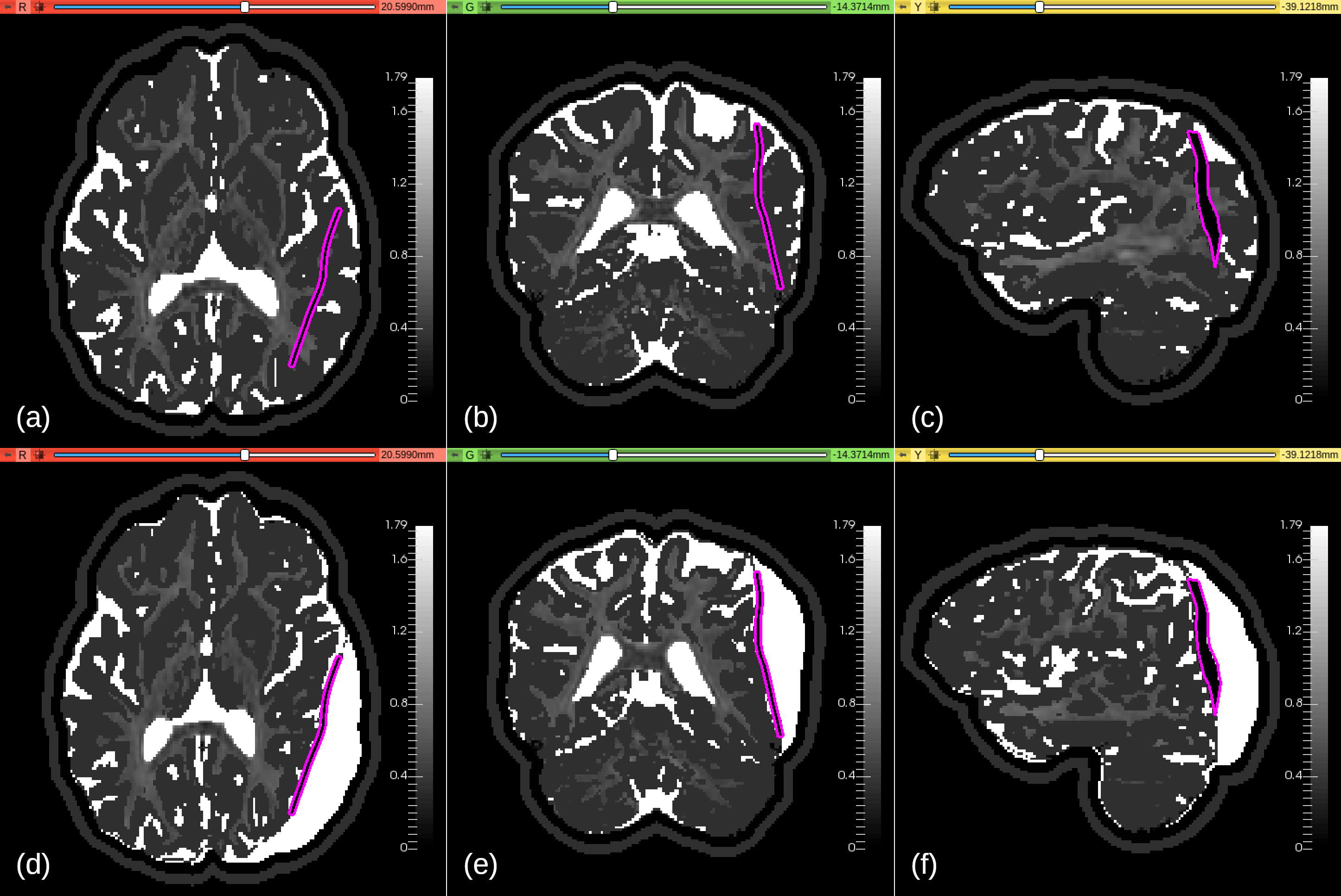}
    \caption{
    Mean conductivity
    (\(\tfrac{1}{3} \mathrm{tr}\,C\))
    for models constructed using
    (a,b,c) original preoperative and
    (d,e,f) deformed by insertion of electrodes postoperative
    image data.
    The ECoG electrode grid substrate is denoted by the purple outline.
    }
    \label{fig:brain-cond}
\end{figure*}

The five tissue classes in the EEG forward model were assigned conductivity tensors as follows.
Isotropic conductivities
(Table~\ref{tab:conductivity})
were assigned to the
scalp,
skull,
cerebrospinal ﬂuid (CSF),
electrode grid array substrate and
gray matter regions
\citep{
  hallez_etal_2007_review,
  vorwerk_etal_2014_guideline}.
The anisotropic conductivity of the white matter
was estimated from the diffusion tensors
using the fractional method
with empirically defined scaling factors
introduced by \citet{tuch_etal_2001_conductivity}.
The mean conductivity
(\(\tfrac{1}{3} \mathrm{tr}\,C\))
of white matter was in the range from
\(1\times10^{-6}\) to 0.99~S/m,%
with an average value of 0.56~S/m,
which agrees with values reported in the literature
\citep{haueisen_etal_1997_influence,tuch_etal_2001_conductivity}.
Fig~\ref{fig:brain-cond} shows the mean conductivity
for the model based on preoperative images
and the model corresponding to the postoperative, deformed configuration of the brain.
The unphysical conductivities of the brain tissue
between the electrode grid array
(represented by the tissue with conductivity close to zero)
and the skull
are clearly visible,
whereas regions further from the implanted electrodes
(and therefore further from the deformed brain surface)
have similar conductivities in both images.

\begin{table*}[t]
  \centering
  \caption{Conductive compartments used in the patient-specific EEG forward models.}
  \label{tab:conductivity}
  \begin{tabularx}{\textwidth}{@{}Xcc@{}}
    \toprule
    Compartment              & Conductivity (S/m) & References \\
    \midrule
    Scalp                    & 0.33               & \citet{geddes_baker_1967_specific,stok_1987_influence} \\
    Skull                    & 0.012              & \citet{hallez_etal_2007_review,gutierrez_etal_2004_estimating} \\ %
    Cerebrospinal ﬂuid (CSF) & 1.79               & \citet{baumann_etal_1997_electrical} \\
    Electrode grid array     & \(10^{-6}\)        &  \\
    Gray matter              & 0.33               & \citet{geddes_baker_1967_specific,stok_1987_influence} \\
    \bottomrule
  \end{tabularx}
\end{table*}

Patient-specific model of brain bioelectric activity consists therefore of
conductivity (tensor \(C\) in Eq.~\eqref{eq:poisson-eq}--\eqref{eq:poisson-nbc}) distribution
depicted in the bottom row of Fig.~\ref{fig:brain-cond}
together with the (deformed) geometry
(\(\Omega\) in Eq.~\eqref{eq:poisson-eq}--\eqref{eq:poisson-nbc})
depicted in the bottom row of Fig.~\ref{fig:mri-ct-elec}.

\subsection{Solution procedure for the EEG forward problem}
\label{sec:solution-eeg}

The finite element method using a structured hexahedral mesh
is an attractive choice
because the mesh can be directly generated from voxel-based medical images,
whereas the generation of surface-based tetrahedral meshes is more complicated
\citep{
  schimpf_etal_1998_realistic,
  schimpf_etal_2002_dipole,     %
  rullmann_etal_2009_eeg,        %
  vorwerk_etal_2017_mixed,
  wolters_etal_2007_geometryadapted}.
Although geometry-adapted hexahedral and tetrahedral meshes can achieve better accuracy
than regular hexahedral meshes with similar number of degrees of freedom
\citep{wolters_etal_2007_geometryadapted,vorwerk_etal_2017_mixed},
it is recognized that creating patient-specific, geometry-conforming meshes
is not feasible in clinical applications
\citep{wittek_etal_2016_finite,vorwerk_etal_2017_mixed}.
The generation of a regular hexahedral mesh takes
advantage of the cubic voxel structure which is inherent to medical images.
This greatly simplifies mesh generation which is an important consideration
for ensuring compatibility with clinical workflows.
Moreover, regular hexahedra have better numerical properties
than skewed hexahedra or tetrahedra
\citep{hughes_2000_finite}.
In this study,
we used a regular hexahedral mesh with resolution of 1~mm,
the same resolution as that used in similar previous studies
\citep{haueisen_etal_2002_influence,rullmann_etal_2009_eeg,vorwerk_etal_2017_mixed},
which ensures that the CSF compartment is appropriately modeled
and the skull is a closed compartment.

The conductivity tensors were assigned directly from voxels
to integration points in the elements.
We refer to this approach as the ``image-as-a-model'' concept
because the finite element mesh used to solve the problem
corresponds directly to the image data
with one-to-one correspondence between mesh elements and image voxels.
This eliminates time-consuming generation of body-fitted meshes
and results in a structured hexahedral mesh with perfect element quality
since all elements are cubes.
The finite element mesh for the EEG forward problem contained
1,618,745 nodal points and
1,565,095 linear hexahedral elements,
compared to
5,898,240 voxels in the original images
(the excluded elements correspond to voxels of air outside the head).

The EEG forward problem solution method was implemented using the open-source MFEM library
\citep{anderson_etal_2020_mfem}
(\url{https://mfem.org}).
A continuous Galerkin formulation
with linear hexahedral elements was used for the spatial discretisation
of the potential and for computing its gradient, the electric field.
The finite element method is based on the weak form of the governing equations
which means that the zero-flux boundary condition is naturally satisfied.
This is in contrast to strong form methods such as the finite difference method
which require more elaborate treatment of the Neumann boundary conditions
\citep{saleheen_ng_1997_new,bourantas_etal_2020_fluxconservative}.

Various approaches for modeling the dipole source have been proposed, including the
partial integration direct potential approach
\citep{yan_etal_1991_finiteelement,schimpf_etal_2002_dipole},
subtraction approach
\citep{vandenbroek_etal_1996_computation,wolters_etal_2007_numerical,drechsler_etal_2009_full},
Whitney elements
\citep{tanzer_etal_2005_representation},
Raviart--Thomas elements
\citep{pursiainen_etal_2012_complete}
and the Saint-Venant direct potential approach
\citep{buchner_etal_1997_inverse,medani_etal_2015_fem}.
In this study, the full subtraction approach
\citep{drechsler_etal_2009_full}
was used to model the current dipole source.
The subtraction approach has been shown to provide accurate results
for realistic 3D problems
\citep{schimpf_etal_2002_dipole,drechsler_etal_2009_full}
although it may not perform as well as a modified Saint-Venant method
for the case of sources close to the interface of layers
with different conductivities
\citep{medani_etal_2015_fem}.

The discretised equations were solved using the conjugate gradient (CG) method
with an algebraic multigrid (AMG) preconditioner
from the HYPRE library of linear solvers
(\url{http://www.llnl.gov/casc/hypre}).
The computation times
for solving the EEG forward problem
and constructing the lead fields
are discussed in the results section.

\subsection{Quantitative measures}
\label{sec:methods:quantitative-measures}

We used the following measures to quantify the differences
in the topology and magnitude
between the predictions made using models based on
the original preoperative and the
deformed by insertion of electrodes
postoperative images.
The relative difference metric (RDM)
is a measure of the difference in the shape of two data sets
and is defined as follows
\citep{meijs_etal_1989_numerical}:
\begin{equation}
    \mathrm{RDM}(\bm{y}, \hat{\bm{y}}) = \sqrt{\sum_{i=1}^n \left (
    \frac{\hat{\bm{y}}_i}{\sqrt{\sum_{i=1}^n \hat{\bm{y}}_i^2}} -
    \frac{    {\bm{y}}_i}{\sqrt{\sum_{i=1}^n     {\bm{y}}_i^2}}
    \right ) }.
\end{equation}
The magnitude factor (MAG)
is a measure of the difference in magnitude
between the two
and is defined as follows
\citep{meijs_etal_1988_relative}:
\begin{equation}
    \mathrm{MAG}(\bm{y}, \hat{\bm{y}}) = \sqrt{\frac{\sum_{i=1}^n {\hat{\bm{y}}_i}^2}{\sum_{i=1}^n {\bm{y}_i}^2}}.
\end{equation}

We compare lead fields between models
by selecting $\bm{y}$ and $\hat{\bm{y}}$ to be stacked vectors
of the three columns of each lead field matrix
(i.e., the three spatial components of the electric field,
\(\bm{E}=-\nabla u\),
for each lead)
associated with a specific spatial location
\citep{hyde_etal_2018_comparison}.
For each comparison between two models,
this resulted in a single RDM and MAG value for each element or voxel.
This provides a spatially varying map of model similarity.
The RDM and MAG metrics measure how the topography and magnitude
of the predicted electrode voltages will vary based on model selection.
The difference between the models is smallest when
RDM is close to 0 and
MAG is close to 1
(or, equivalently, \(\mathrm{log}_{10}(\mathrm{MAG})\) is close to 0).

\section{Results}

\subsection{Simulation of electric field originating at seizure onset zone}

As a rough approximation of an epileptic seizure,
we simulated the electric field originating from
a dipole in the seizure onset zone (SOZ).
This example serves to demonstrate the application of the proposed methodology,
and to assess the effect of the change in model geometry and conductivity
on the EEG forward model predictions.

To evaluate the effect of the model geometry on the predicted
electric potential within the brain and on the surface electrodes,
we solved the iEEG forward problem with a current dipole source using
the original image data with both the actual and projected electrode locations
(Fig.~\ref{fig:brain-cond}, top row),
and
the deformed image data with the actual electrode locations
(Fig.~\ref{fig:brain-cond}, bottom row).
To mimic a current dipole set up by cortical neurons,
we placed a dipole,
with dipole moment of 100~µAmm,
in the gray matter of the brain
at a distance of 13.1~mm to the nearest ECoG electrode.
The dipole
was assumed to be located within the
temporal lobe,
which is a region that is commonly implicated in epilepsy seizure onset localization
\citep{salami_etal_2020_seizure}.

\begin{figure*}
    \centering
    \includegraphics[width=\textwidth]{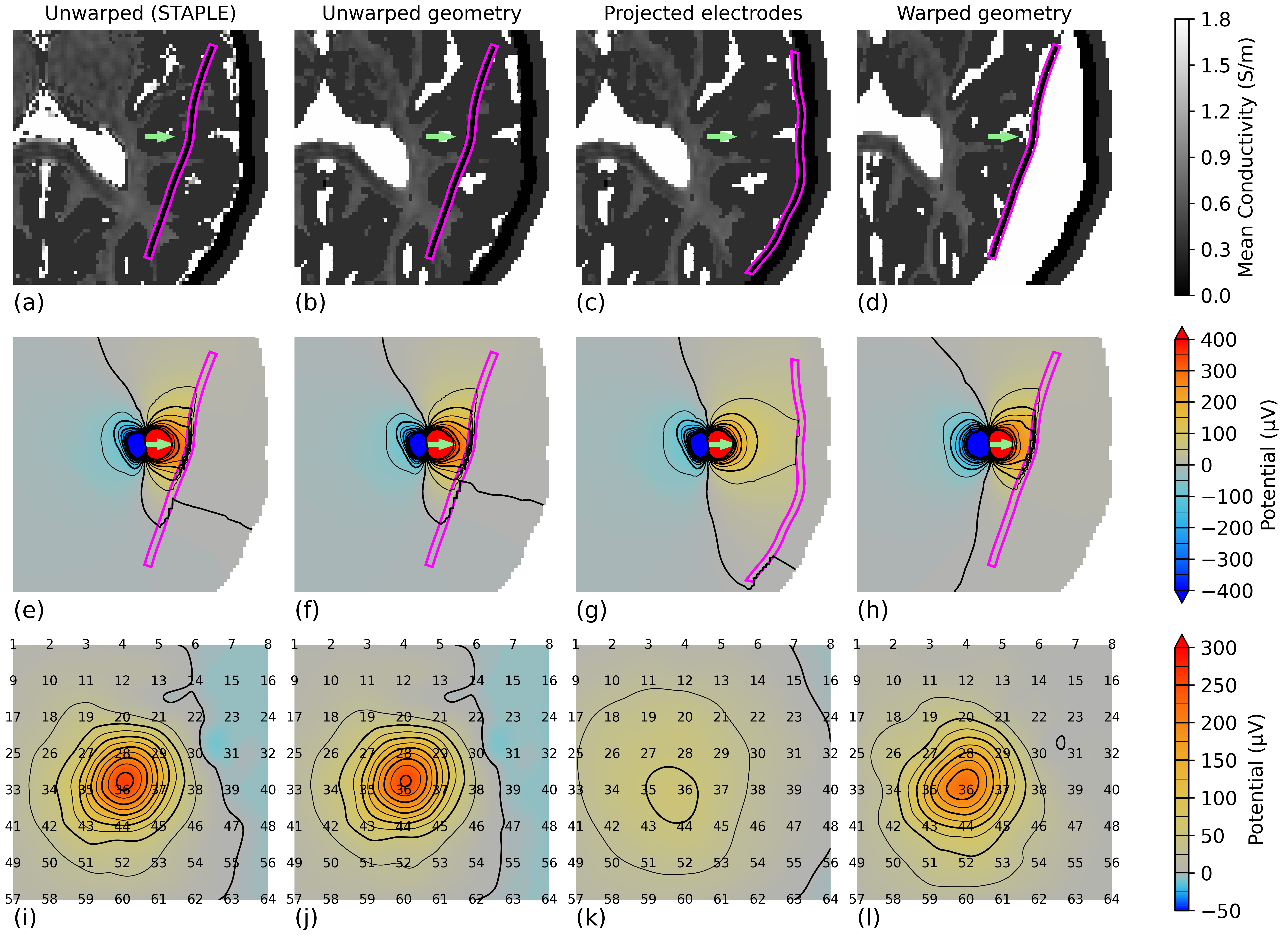}
    \caption{
    Electric potential in the brain
    generated by a current dipole
    as predicted by,
    from left to right,
    the undeformed model based on original preoperative image data
    segmented using the reference STAPLE method and our DTI-based method
    with actual electrode positions (as seen on CT),
    the undeformed model based on original preoperative image data
    segmented using our DTI-based method
    with projected electrode positions,
    and the model based on image data deformed by implantation of electrodes
    with actual electrode positions.
    Top and middle rows:
    axial slices of the brain
    (in a plane coincident with the dipole)
    showing
    (a, b, c, d) mean conductivity and
    (e, f, g, h) predicted electric potential.
    The current dipole moment vector is denoted by the green arrow and
    the ECoG electrode grid substrate is denoted by the purple outline.
    Bottom row:
    (i, j, k, l) predicted electric potential
    on ECoG electrode grid.
    }
    \label{fig:brain-dipole-potential}
\end{figure*}

Fig.~\ref{fig:brain-dipole-potential}
shows the distribution of electric potential within the brain
predicted by the preoperative models based on the original image data
segmented using the reference STAPLE method and our DTI-based method
(with actual and projected electrode positions), and
the postoperative model based on the deformed image data
(also segmented using our DTI-based method
and with actual electrode positions).

\begin{figure*}
    \centering
    \includegraphics[width=\textwidth]{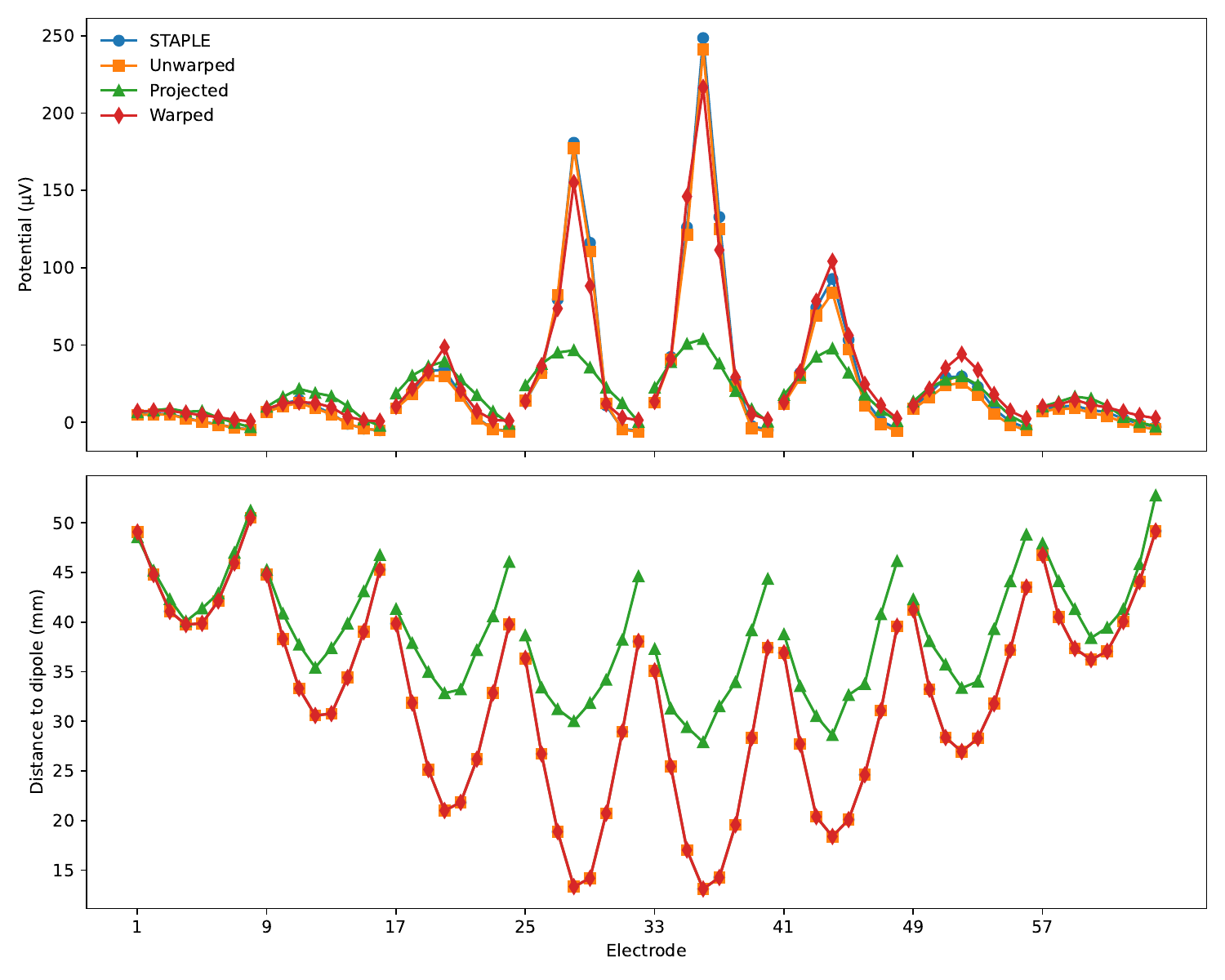}
    \caption{
    Electric potential at ECoG brain surface electrodes
    generated by a current dipole
    as predicted by
    the unwarped
    (constructed using the original preoperative image data
    segmented using the STAPLE reference method and our DTI-based method
    with actual electrode positions),
    the projected
    (constructed using the original preoperative image data
    with projected electrode positions), and
    the warped
    (constructed using the deformed image data
    that corresponds to the predicted postoperative configuration of the brain)
    iEEG forward models.
    }
    \label{fig:brain-dipole-barchart}
\end{figure*}

There appears to be only a small difference in electric potential
predicted by the preoperative models with unwarped geometry
created using the STAPLE and DTI-based segmentations
(Fig.~\ref{fig:brain-dipole-potential}).
The difference in potential at the 64 electrodes
predicted by the STAPLE and DTI-based segmentations is negligible
(Fig.~\ref{fig:brain-dipole-barchart}),
with RDM of 0.04 and MAG of 0.96.
The relatively small difference between these two models is to be expected
because the underlying brain geometry and electrode positions
are the same in both models.
The small difference in the results may be explained by the
localized differences in the conductivity distribution
obtained by the two different segmentation methods
with some voxels belonging to different tissue classes in each model.

The difference in the topography and magnitude
of the electric potential predicted by
the preoperative (unwarped geometry) and
the postoperative (projected electrodes and warped geometry) models
appears to be significant.
Moreover, there is a large difference in the electric potential at the electrodes
predicted by the unwarped, projected and deformed models
(Fig.~\ref{fig:brain-dipole-barchart}).
The difference between the potential at the 64 electrodes computed using
the original image data with the actual electrode positions and
the deformed image data
is relatively large with
RDM of 0.19 and MAG of 1.02.
The difference in electric potential predicted by the unwarped and warped models
is significant and may be explained by the large differences in conductivity distribution
due to the brain shift caused by implantation of electrodes.
These results suggest that the incorporation of brain shift in the EEG forward model
may affect the source localization significantly.
The difference between the potential at the 64 electrodes computed using
the original image data with the projected electrode positions and
the deformed image data
is significant with
RDM of 0.53 and MAG of 0.48.
Here, the difference can be explained in part by the increased distance
between the dipole and the electrodes in the projected electrode model,
in addition to the difference in the underlying brain geometry as discussed above.

Our results for this example suggest that the modeling error
introduced by using incorrect tissue geometry
obtained from the original preoperative images
(instead of the deformed images
that correspond to the postoperative configuration
of the brain with implanted electrodes)
is significant and we expect this to affect the accuracy of source localization.
Conversely,
we expect that significant improvements in source localization accuracy may be realized
by applying the modeling strategies proposed in this study.

\subsection{Towards the solution of the inverse problem: lead field matrices}

One of the most important techniques
for measurement of brain activity,
especially in the treatment of epilepsy and brain tumors,
is source localization using EEG
\citep{grech_etal_2008_review,brette_destexhe_2012_handbook}.
Source localization requires the solution of the EEG inverse problem
which involves locating the current source
given sparse data of the electric potential from electrode recordings.
Many inverse solution methods rely on the lead field matrix
which describes the sensitivity patterns of the EEG sensors
\citep{weinstein_etal_2000_leadfield,grech_etal_2008_review,brette_destexhe_2012_handbook}.
The lead field matrix can be constructed
by solving the EEG forward problem
for each ground and sensor electrode pair.
The lead field values are computed using reciprocity
as the gradient of the computed potential
for each ground--sensor electrode pair forward problem,
which produces three lead field columns per spatial location.
Once the lead field matrix has been computed
it can be used to calculate the potential at the electrodes
produced by a dipole located within any element of the model.

To assess the effect of the geometry and conductivity distribution
on the expected source localization accuracy,
we compared lead field matrices computed
using the models based on the original
(with actual and projected electrode locations,
and segmented using the STAPLE and DTI-based methods)
and deformed image data.
To construct the lead fields,
the first electrode was chosen arbitrarily as the ground electrode,
and a unit current source was applied to each of the remaining 63 electrodes.
The dimensions of each lead field matrix was
\( (64-1) \times (3 \times 1,565,095) = 63 \times 4,695,285\).

Computations were performed on a single core of a laptop computer
with Intel Core i7-8750H 4.10~GHz CPU and 32~GB RAM.
The total computation time
for solving 63 forward problems
to construct a lead field matrix for 64 electrodes
was less than 15 minutes.
Assembly of the stiffness matrix
(which can be precomputed and reused for different load cases)
took around 1~min,
and each row of the lead field matrix was solved within 13~s
(including imposition of boundary conditions,
solution of the linear system of equations,
and recovery of the gradient of the solution at the element centroids).

To quantify the difference between the lead fields,
the RDM and MAG metrics (section~\ref{sec:methods:quantitative-measures})
were applied to compare the three corresponding columns from lead field matrices
from two different volume conductor models
at each spatial location.
The values in the three corresponding columns of a lead field matrix
are the sensitivities of all electrode voltage measurements
to the presence of electrical current at a particular point
(voxel or element)
within the head.
For each comparison between two models,
this resulted in a single RDM and MAG value for each voxel.

\begin{figure*}
    \centering
    \includegraphics[width=\textwidth]{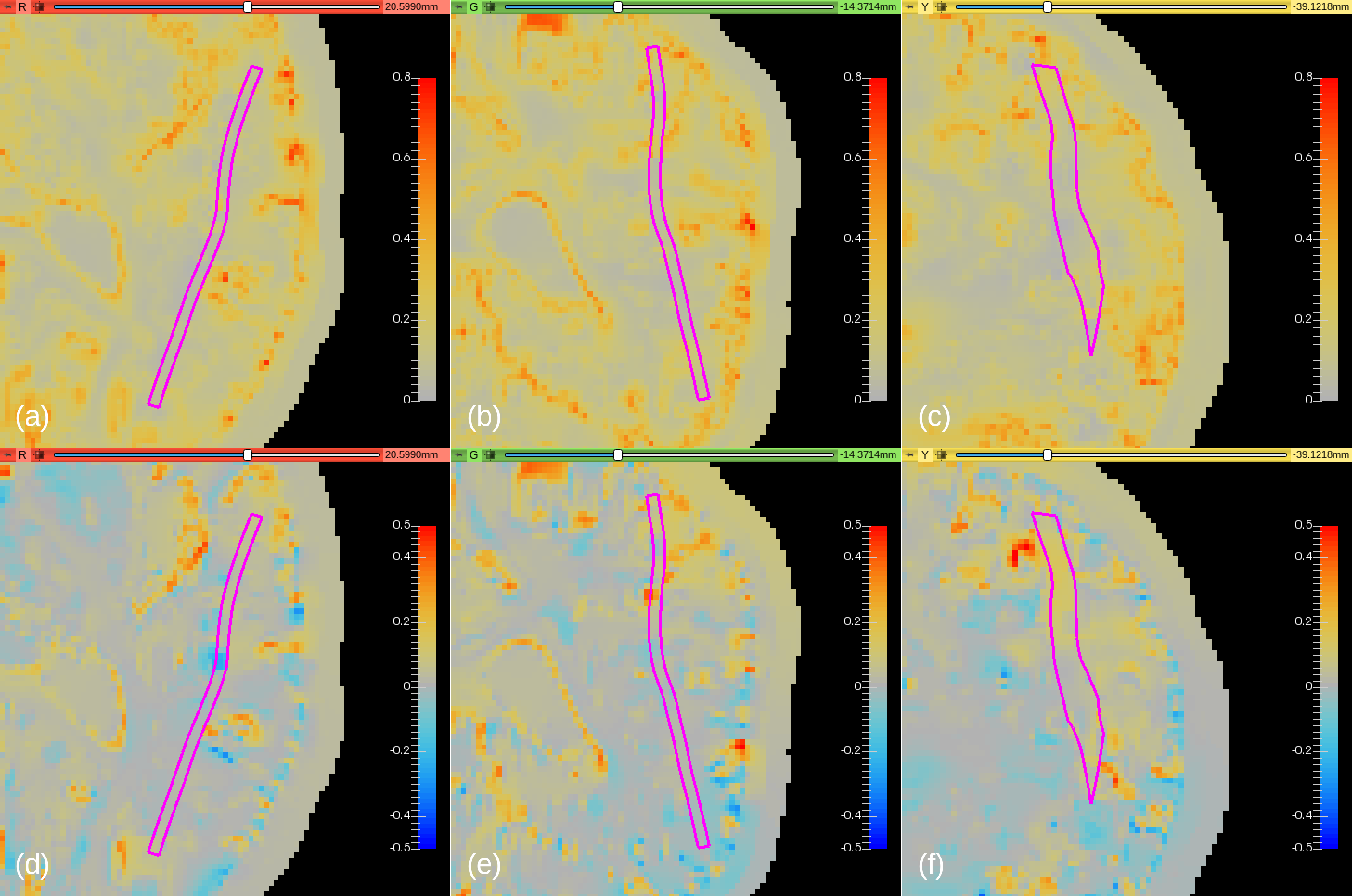}
    \caption{
    Difference in lead fields computed using
    the preoperative model
    with actual electrode positions as seen on CT
    created using segmentations obtained by the STAPLE method
    and the proposed DTI-based tissue classification method.
    Top row: topographic difference (RDM)
    in (a) axial, (b) coronal and (c) saggital slices of the brain.
    Bottom row: magnitude difference (\(\mathrm{log}_{10}(\mathrm{MAG})\))
    in (d) axial, (e) coronal and (f) saggital slices of the brain.
    The ECoG electrode grid substrate is denoted by the purple outline.
    }
    \label{fig:unwarpedstaple-leadfield-images}
\end{figure*}

Fig.~\ref{fig:unwarpedstaple-leadfield-images}
shows the difference between the lead fields
computed using the original image data
segmented using the STAPLE and DTI-based methods.
There are small differences in the lead fields
especially in the regions at the interface between CSF and brain tissue
where voxels are classified differently by the two segmentation methods.
The differences between the lead fields
computed using the two segmentation methods
are small compared to those between the lead fields
computed using preoperative (unwarped geometry)
and postoperative (projected electrodes and warped geometry) models
as discussed below.
This suggests that small errors in segmentation will have a negligible effect
compared to the change in geometry caused by brain shift.

\begin{figure*}
    \centering
    \includegraphics[width=\textwidth]{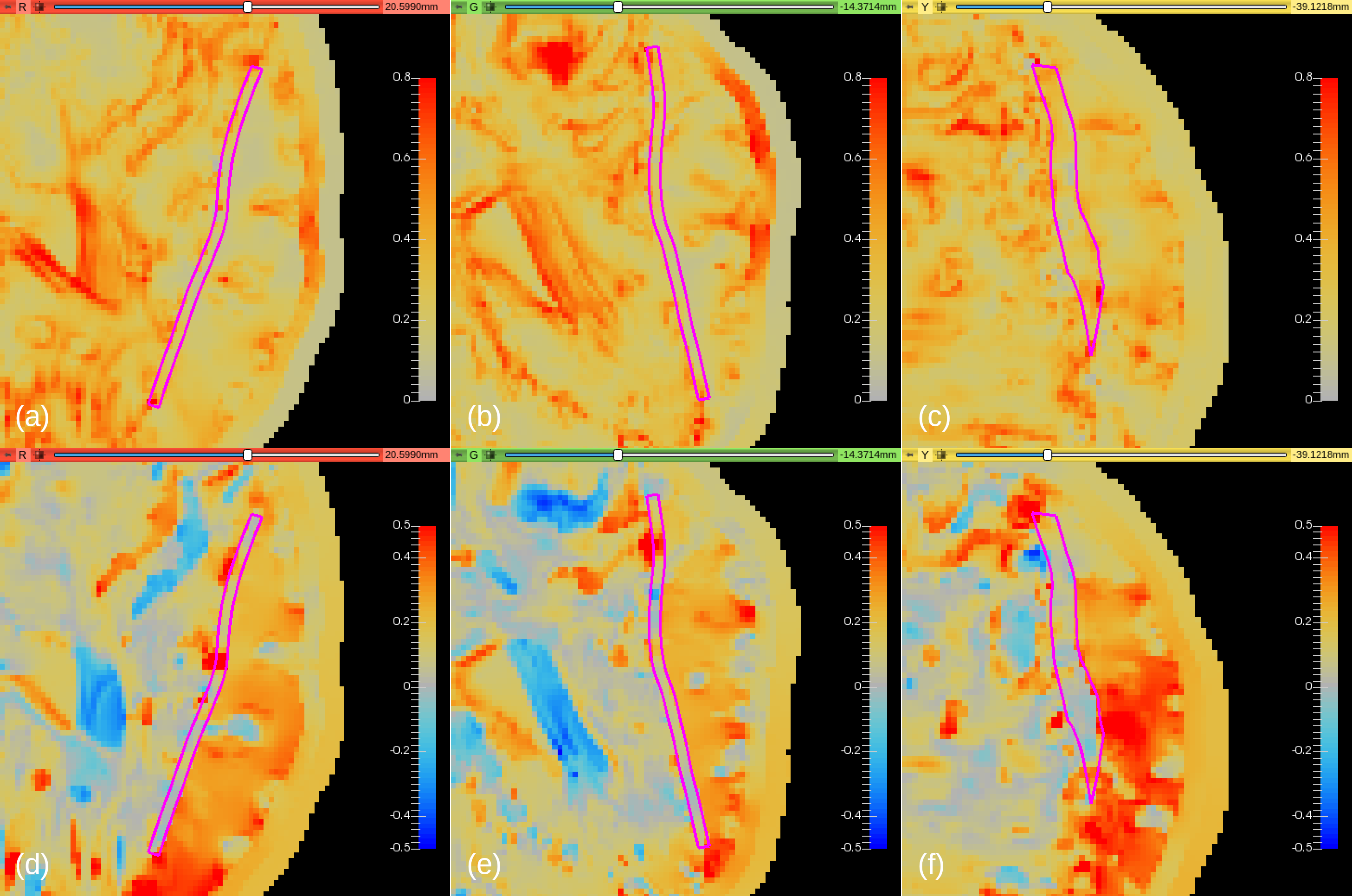}
    \caption{
    Difference in lead fields computed using
    the preoperative model
    with actual electrode positions (as seen on CT)
    and
    the deformed by insertion of the electrodes postoperative model
    (also with actual electrode positions).
    Top row: topographic difference (RDM)
    in (a) axial, (b) coronal and (c) saggital slices of the brain.
    Bottom row: magnitude difference (\(\mathrm{log}_{10}(\mathrm{MAG})\))
    in (d) axial, (e) coronal and (f) saggital slices of the brain.
    The ECoG electrode grid substrate is denoted by the purple outline.
    }
    \label{fig:unwarped-leadfield-images}
\end{figure*}

\begin{figure*}
    \centering
    \includegraphics[width=\textwidth]{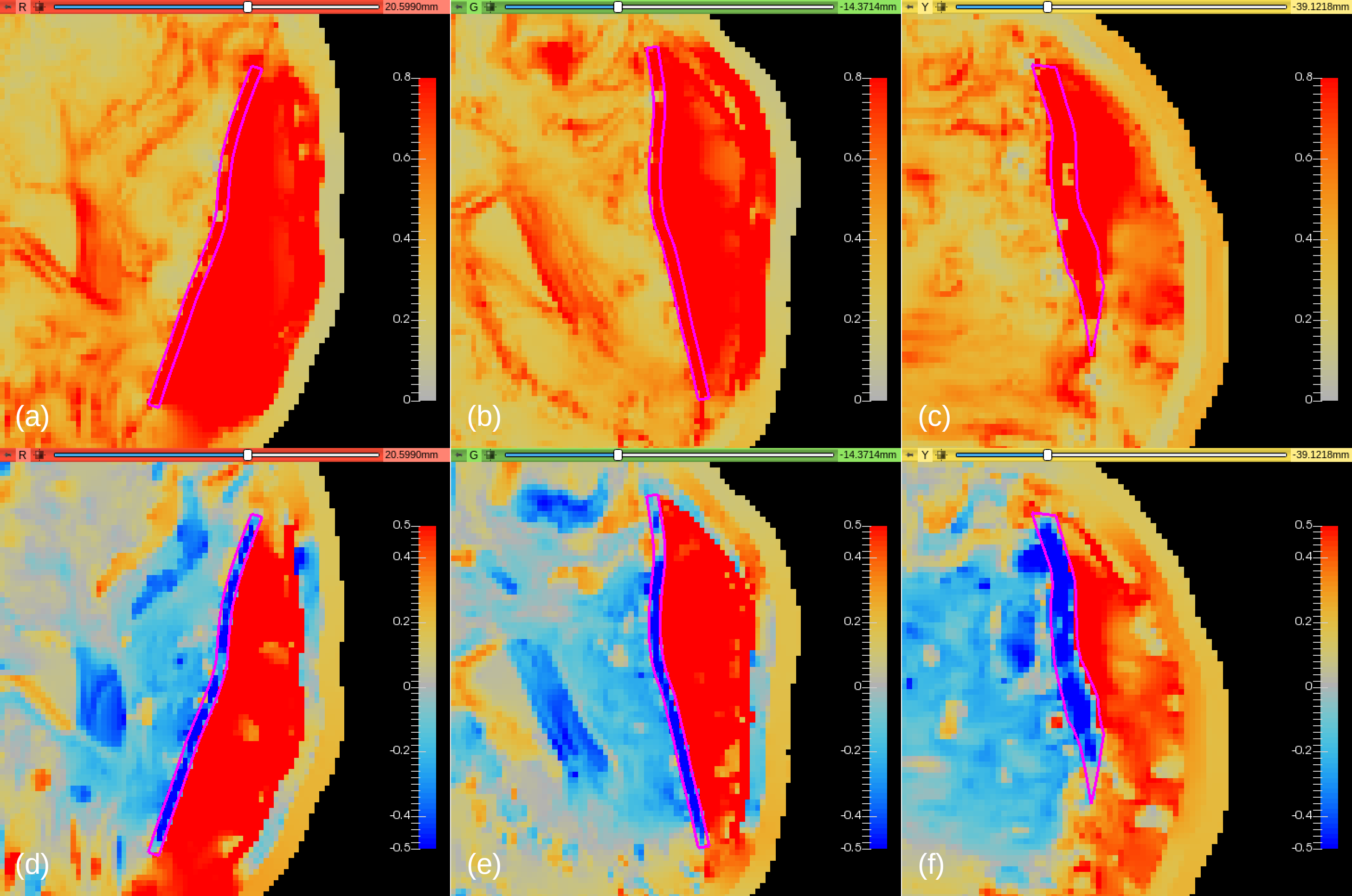}
    \caption{
    Difference in lead fields computed using
    the preoperative model
    with electrode positions projected onto the brain surface
    and
    the deformed by insertion of the electrodes postoperative model
    with actual electrode positions (as seen on CT).
    Top row: topographic difference (RDM)
    in (a) axial, (b) coronal and (c) saggital slices of the brain.
    Bottom row: magnitude difference (\(\mathrm{log}_{10}(\mathrm{MAG})\))
    in (d) axial, (e) coronal and (f) saggital slices of the brain.
    The ECoG electrode grid substrate is denoted by the purple outline.
    }
    \label{fig:projected-leadfield-images}
\end{figure*}

Figs.~\ref{fig:unwarped-leadfield-images} and~\ref{fig:projected-leadfield-images}
show the differences between the lead fields
computed using the original image data,
with actual and projected electrode locations, respectively,
and the deformed image data.
The results show that the difference in the lead fields is greatest
in the region close to the electrode grid array.
This is to be expected because the region closest to the electrodes
corresponds with the greatest amount of tissue deformation.
There are also significant differences within the same hemisphere.
The electrodes are usually placed close to the expected source location
which means that the greatest differences coincide with the region of the brain
that is most likely to contain the seizure onset zone.
This can be expected to have a detrimental effect
on the accuracy of source localization.

The most significant effect on the differences in the lead field matrices
(in both the RDM and MAG)
appears to be the misclassification of brain tissue as CSF,
and vice versa,
in the model that is based on the original preoperative MRI.
In these regions
the RDM is greater than 0.5,
and MAG is less than 1/3
(underestimated by a factor of 3)
or greater than 3
(overestimated by a factor of 3).
This is in line with studies on model sensitivity to conductivity
which reported that while small uncertainties in the CSF conductivity
have a negligible effect on the result of dipole reconstruction,
outright incorrect tissue classification
has strong effects on the forward solutions
\citep{vorwerk_etal_2014_guideline,vorwerk_etal_2019_influence}.

The comparison between the lead fields
computed using the original (actual preoperative),
with both the actual and projected electrode positions, and
the deformed (predicted postoperative) image data
suggest that accurate source localization
based on iEEG or ECoG recordings
requires accurate classification of tissue
in the postoperative configuration of the brain,
after the electrodes have been implanted.

\section{Discussion}

Intracranial electroencephalography (iEEG) and electrocorticography (ECoG)
are in clinical practice often used without numerical modeling
\citep{khosropanah_etal_2020_eeg,
  ryvlin_etal_2014_epilepsy,
  scherg_etal_2019_taking},
which can certainly be considered useful in planning of epilepsy surgery.
Neverthelss, these methods are qualitative, and, even when successful,
may lead to large resections
\citep{vakharia_etal_2018_getting}.
Advanced mathematical modeling and analysis of iEEG and ECoG signals
may one day lead to single voxel size accuracy in SOZ localization.
In this paper,
we presented a novel methodology for patient-specific solutions of
the iEEG or ECoG forward problem
that accounts for the brain shift caused by craniotomy
and insertion of subdural grid electrodes.
The method relies on biomechanics-based image warping
to transform the original preoperative image data
to the predicted postoperative configuration
with implanted intracranial electrodes.
Through the analysis of a real, patient-speciﬁc
case from Boston Children’s Hospital,
we have shown that it is possible to apply biomechanical modeling
using a meshless approach to compute the deformation
field within the brain arising from invasive electrode placement.
We used the computed deformation field to warp preoperative MRI and DTI
into the postoperative conﬁguration of the brain.
This provides a highly detailed map of the electrodes relative to neurological landmarks,
and an accurate representation of the postoperative brain geometry
which is required as an input to the iEEG forward model.

We generated an efficient pipeline to numerically solve the iEEG forward problem
on real patient-specific data, that consists of a fast classification algorithm
using diffusion tensor images.
The automated DTI-based brain tissue classification takes less than a minute.
The time required to solve the biomechanical model for image warping
was approximately 10~min,
and a further hour to process the results and register the deformation field.
Once the deformed geometry is made available,
the construction of the iEEG forward model can be completed
in a few hours.
The solution of the iEEG forward problem takes a few minutes,
which includes assembly of the system of linear equations and its solution,
as well as data input and output.

The generation of the patient-specific computational model for the iEEG forward problem,
including
skull stripping,
segmentation of electrodes,
biomechanics-based image warping,
brain tissue classification,
conductivity tensor estimation,
fusion of preoperative and deformed image data
and generation of the finite element mesh,
took an experienced analyst a total of approximately two days.
With further refinement of the modeling pipeline
and accumulated experience from additional cases
we expect this time to be reduced to about 4 hours per patient.
This is acceptable in the research environment
and could be considered to be sufficient for clinical applications.
As close to real-time processing speeds are not demanded by this application,
these simulation and analysis times are compatible with existing clinical workflows.
These timeframes would easily fit within the period of 5--7 days of data collection,
while the electrodes are in the brain.

Most of the modeling steps have been automated
in anticipation of their application in a clinical environment.
The complexity of the preprocessing steps was simplified
through the use of an image-based approach
that circumvents traditional segmentation and meshing.
The iEEG forward model is composed of hexahedral elements
that match the image geometry with one-to-one correspondence between voxels and elements.
This is the highest resolution that can be attained
and should provide maximally accurate simulation results
as better patient-specific data than the voxelized image is not available.
We refer to this as the image-as-a-model approach
because the computational grid is created directly from the image data.

To demonstrate the applicability of the proposed approach
we applied our methodology to a representative epilepsy case
and solved two relevant example problems.
The models of the brain with a current dipole
showed large differences in the electric potential predicted using
the original (actual preoperative) image data and
the deformed (predicted postoperative) image data.
The lead field matrices,
typically used for source localization,
computed using the different models
also showed significant differences.
The results show that the tissue geometry and conductivity has a significant influence on the results
which suggests that significant improvements in source localization accuracy
may be realized by applying the methods described in this study.

Despite the demonstrated efficiency and accuracy of the proposed method
there are a few shortcomings that should be addressed in future studies.
One of the main difficulties of the proposed approach
is the reconstruction of the brain geometry and tissue conductivity maps
from the image data.
The accuracy of the tissue classification
is a limitation that affects the accuracy of the model geometry
and the conductivity distribution.
Although the method does not require connected segments
and surface meshes extracted from these segments,
it does require accurate classification of the tissue type
of each voxel within the brain.
Accurate classification of voxels often requires manual segmentation
but this is a difficult and subjective process that does not guarantee repeatability.
Automated tissue classification methods appear as the most promising avenue
for reducing the effort required to produce accurate label maps of the brain
while simultaneously eliminating variability inherent in manual segmentations.
The segmentation method based on the DTI that we used in this study is relatively simple
but demonstrates the potential to fully automate the tissue classification process.
More advanced automated segmentation procedures have been proposed
\citep{wen_etal_2013_brain}
and these should be considered if higher accuracy is required.
Improvements in the segmentation and tissue classification procedures
are expected to improve the solution accuracy of the iEEG forward problem.
Finally, we considered only a single patient case in this study.
To fully evaluate the methods more patient cases are needed.
The patient case analyzed in this study
can be considered as a proof-of-concept
that demonstrates the accuracy and efficiency of the proposed approach.

In this study,
we used an automated DTI-based tissue classification method
to classify CSF, white matter and gray matter.
However, the availability of DWI is not always guaranteed.
In this case, the conventional MRI (instead of DTI) may be transformed
to the post-implantation configuration
using the same methods described herein,
and the DTI-based method may be replaced
with traditional segmentation methods based on MRI,
such as those available in FreeSurfer
(\url{http://surfer.nmr.mgh.harvard.edu})
\citep{dale_etal_1999_cortical}.
Homogeneous isotropic conductivity would then be assigned to the white matter
because anisotropic conductivity assignment also relies on DWI.
Alternatively, traditional MRI-based segmentation
may be combined with DTI-based anisotropic tissue conductivity assignment.
The choice of segmentation method and tissue conductivity assignment
will typically be based on the accuracy requirements,
and the time constraints in the research or clinical settings.

A tangential area of interest in the surgical treatment of epilepsy
is the modeling of depth electrodes
that are inserted stereotactically through openings in the skull.
Unlike the grid array electrodes considered in the current study,
depth electrodes are long and slender needles
that are inserted deep into the brain parenchyma.
The electrodes are difficult to control and locate during their insertion,
and may cause deformation of the surrounding tissues.
Biomechanics-based image registration as described in this paper,
combined with a suitable needle insertion model
\citep{wittek_etal_2020_mathematical},
may provide useful guidance for surgeons during these procedures.

Our proposed approach for generating patient-specific iEEG or ECoG forward models
for epilepsy patients with implanted subdural electrode grids
has significant potential for clinical applications.
Results show that the model based on the predicted postoperative image data
obtained using biomechanics-based image warping
produces lead fields with significant differences
from those computed using the original image data,
especially in the regions that are close to the electrodes
and areas typically implicated in epileptic seizures
that are of high interest to the epileptologist.
The computation time required to solve the EEG forward problem
is short enough to make the solution of the EEG inverse problem feasible.
We are currently working on automating more of the modeling tasks
to further reduce the time required to construct patient-specific models.
Combining the modeling pipeline described in this paper
with a suitable method for solving the EEG inverse problem
will enable accurate source localization for epilepsy patients
who have undergone invasive electrophysiological monitoring
and this will be the focus of our future investigations.

\section*{Acknowledgments}

A. Wittek and K. Miller acknowledge the support by the Australian Government through
National Health and Medical Research Project Grant no.\ APP1162030.

\bibliography{manuscript.bib}

\nolinenumbers

\newpage
\listoffixmes

\end{document}